\documentclass[aps,showpacs,eqsecnum,twocolumn]{revtex4-1}
\usepackage[english]{babel}
\usepackage{graphicx}
\usepackage{stmaryrd}
\usepackage{amsmath, amsfonts}
\usepackage{mathrsfs}
\usepackage{braket}
\usepackage{color}
\usepackage{xspace}
\usepackage{amscd}
\usepackage{bm}
\definecolor{lightblue}{rgb}{0.13, 0.26, 0.99}
\definecolor{strawberry}{RGB}{229, 0, 49}

\usepackage[
colorlinks=true,
urlcolor=blue,
citecolor=blue,
linkcolor=blue,
hyperfootnotes=false]{hyperref}

\usepackage{longtable}

\newcommand{\e}{\varepsilon}

\newcommand{\na}{\mathrm{NaCuMoO_4(OH)}}

\allowdisplaybreaks

\begin{document}
\title{Angular dependence of electron spin resonance for detecting quadrupolar liquid state of frustrated spin chains}
\author{Shunsuke C. Furuya}
\affiliation{Condensed Matter Theory Laboratory, RIKEN, Wako, Saitama 351-0198, Japan}
\date{\today}
 \begin{abstract}
  Spin nematic phase is a phase of frustrated quantum magnets with a quadrupolar order of electron spins.
  Since the spin nematic order is usually masked in experimentally accessible quantities,
  it is important to develop a methodology for detecting the spin nematic order experimentally.
  In this paper we propose a convenient method for detecting quasi-long-range spin nematic correlations
  of a quadrupolar Tomonaga-Luttinger liquid state of $S=1/2$ frustrated ferromagnetic spin chain compounds, 
  using electron spin resonance (ESR).
  We focus on linewidth of a so-called paramagnetic resonance peak in ESR absorption spectrum.
  We show that a characteristic angular dependence of the linewidth on the direction of magnetic field arises in
  the spin nematic phase.  
  Measurments of the angular dependence give a signature of the quadrupolar Tomonaga-Luttinger liquid state.
  In our method we change only the direction of the magnetic field, keeping the magnitude of the magnetic field and the temperature.
  Therefore, our method is advantageous for investigating the one-dimensional quadrupolar liquid phase 
  that usually occupies only a narrow region of the phase diagram.
 \end{abstract}
\maketitle

\section{Introduction}\label{sec:intro}

\begin{table*}[t!]
 \begin{tabular}{ccc}
  \hline\hline
  anisotropy & standard TLL & quadrupolar TLL  \\ 
  \hline
  intrachain exchange interaction  \eqref{H'_ex}  & $1+\cos^2\theta$ & $\color{strawberry}{\sin^2\theta\cos^2\theta}$  \\
  staggered DM interaction \eqref{H'_DM} & $\sin^2\theta$ & $\color{strawberry}{\sin^2\theta\cos^2\theta}$ \\
  unfrustrated interchain exchange interaction \eqref{H'_uf} & $2\cos^4\theta+\sin^2\theta$ & $\color{strawberry}{\sin^2\theta \cos^2\theta}$  \\ 
  frustrated interchain exchange interaction \eqref{H'_f} & $\cos^4\theta+\cos^2\theta$ & $\color{strawberry}{\sin^2\theta \cos^2\theta}$  \\ 
  \hline \hline
 \end{tabular}
 \caption{Angular dependences of the linewidth of the paramagnetic resonance peak for the standard TLL and the quadrupolar TLL
 when the temperature $T$ is lower than the single-magnon gap \eqref{cond_T} 
 and the magnetic field and the magnetization are weak \eqref{cond_M_relaxed}.
 $\theta$ is the angle between the direction of the magnetic field and 
 a direction of an anisotropy.
 }
 \label{table:width}
\end{table*}

Spin nematic phase is an intriguing phase of quantum magnets characterized by the presence of spontaneous quadrupolar order
and by the absence of spontaneous dipolar order.
It arises out of interplay between geometrical frustration and interaction effects of electron spins.
Geometrical frustration obstructs growth of the spontaneous dipolar order and
the interaction effect facilitates growth of the quadrupolar order.
An attractive interaction is necessary for magnons to form a pair and
to condense prior to a single-magnon condensation~\cite{Shannon_SQ_nematic, Kecke_multimagnon, Vekua_1dnematic,
Sudan_1dnematic, Zhitomirsky_1dnematic}.
As a natural but nontrivial phenomenon, the spin nematic phase has been actively investigated.

Many models are known to exhibit spin nematic 
phases~\cite{Shannon_SQ_nematic, Zhitomirsky_pyrochlore, Momoi_nematic_triangle, Zhitomirsky_1dnematic}.
In particular, an $S=1/2$ frustrated ferromagnetic spin chain is of great
interest~\cite{Chubukov_1dnematic, Zhitomirsky_1dnematic, Kecke_multimagnon, Hikihara_chiral_nematic, Sato_quasi1dnematic, Starykh_1dnematic, Onishi_1dnematic}.
The spin nematic phase of the $S=1/2$ frustrated ferromagnetic chain can be seen as a quadrupolar Tomonaga-Luttinger liquid (TLL) 
phase~\cite{Vekua_1dnematic, Hikihara_chiral_nematic, Sato_1dnematic_NMR, Sato_1dnematic_NMR_2011}.
While a ``standard'' TLL phase of antiferromagnetic spin chains~\cite{Giamarchi_book}
is accompanied by a quasi-long-range dipolar antiferromagnetic order~\cite{Furuya_unionjack},
the quadrupolar TLL phase is accompanied by a quasi-long-range spin nematic order.
The $S=1/2$ frustrated ferromagnetic chain also draws much attention for a simple experimental realization in edge-sharing CuO$_2$ chains.
Thanks to these features, many $S=1/2$ frustrated ferromagnetic chain compounds have been synthesized and investigated until 
today~\cite{Hase_1dnematic_Rb, Masuda_LiCu2O2, Enderle_LiCuVO4, Nawa_1dnematic_Na, Mourigal_LiCuVO4, Nawa_LiCuVO4_NMR, Buttgen_LiCuVO4}.

\begin{figure}[t!]
 \centering
 \includegraphics[bb = 0 0 1300 600, width=\linewidth]{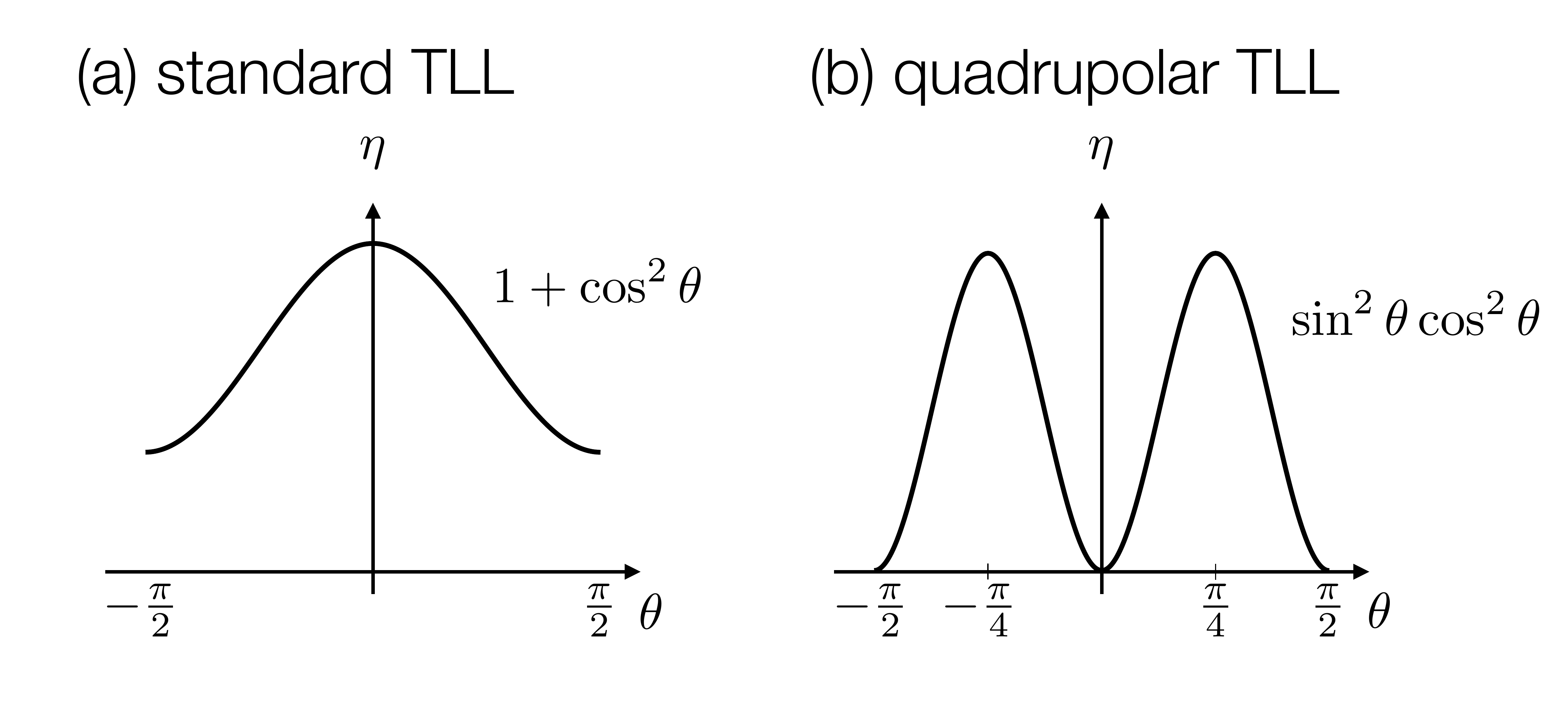}
 \caption{The angular dependence of the ESR linewidth of (a) the standard TLL and (b) the quadrupolar TLL
 induced by an intrachain exchange anisotropy.
 The angle $\theta$ specifies direction of magnetic field.
 At $\theta=0$, the magnetic field is parallel to direction of an anisotropic spin-spin interaction.
 The periodicity of the maximum linewidth against $\theta$ enables us to distinguish the quadrupolar TLL from the standard one.
 }
 \label{width_comparison}
\end{figure}

Nevertheless,
there remains an issue of how to detect experimentally the quasi-long-range nematic order.
Basically experimental techniques are sensitive only to dipolar correlations $\braket{S_{\bm r}^aS_{\bm 0}^a}$
and not to quadrupolar correlations $\braket{S_{\bm r}^aS^b_{\bm r+\bm e}S_{\bm 0}^aS^b_{\bm e}}$.
Several theoretical proposals were made to solve the issue.
A power-law temperature dependence of 
the nuclear magnetic resonance (NMR) relaxation rate $1/T_1\propto T^{2K-1}$ 
gives a signature of the spin nematic phase,
where $K$ is a field-dependent TLL parameter~\cite{Sato_1dnematic_NMR, Sato_1dnematic_NMR_2011}.
Unfortunately, the predicted power law is not yet observed in experiments~\cite{Grafe_1dnematic} probably
because the 1D spin nematic phase appears only in a narrow temperature range.
Changing the temperature, we easily go out of the ideally 1D region of the spin nematic phase.
It was also pointed out recently that the resonant inelastic x-ray scattering method can detect a quadrupolar operator~\cite{Savary_RIXS}.
This proposal is yet to be examined experimentally.

In this paper we propose a practical way of detecting quasi-long-range spin nematic correlations of $S=1/2$ frustrated ferromagnetic chain compounds.
It is to investigate dependence of linewidth of an electron spin resonance (ESR) absorption peak on the direction of magnetic field.
We point out that the linewidth is sensitive to nematic correlations of electron spins.
As a result of the sensitivity, our method gives a qualitative characterization of the quadrupolar TLL.
The main result is summarized in Fig.~\ref{width_comparison} and Table~\ref{table:width}.
Changing the field direction on a plane, we can distinguish the quadrupolar TLL from the standard TLL.
They are distinguished by a period of an angle $\theta$ of the magnetic field that maximizes the linewidth.
The linewidth of the standard TLL becomes maximum at $\theta=0$ or $\pi/2 \mod \pi$.
The angle $\theta$ that maximizes the linewidth depends on anisotropies (Table~\ref{table:width}) and also on the definition of $\theta$.
However, in any case, the period is $\pi$.
In contrast to the standard TLL, 
the linewidth of the quadrupolar TLL becomes maximum at intermediate angles $\theta=\pi/4\mod \pi/2$ (Table~\ref{table:width}).
The period is $\pi/2$.
This distinction is effective 
(1) when the temperature is lower than a single-magnon excitation gap~\cite{Sato_quasi1dnematic, Onishi_1dnematic},
(2) when a weak anisotropic exchange interaction and/or a weak staggered Dzyaloshinskii-Moriya (DM) interaction are present,
and (3) when the magnetic field is weak compared to temperature.
The condition on the temperature is necessary to rule out effects of gapped single-magnon excitations
and that on the magnetic field is to make the linewidth finite.

This paper is planned as follows.
In Sec.~\ref{sec:quad} we review the quadrupolar TLL phase of the $S=1/2$ frustrated ferromagnetic chain.
Section~\ref{sec:esr} is an introduction to ESR of quantum spin systems,
where we will get a glimpse of the way to detect nematic correlations through the main peak of the ESR spectrum.
We call in this paper the main peak a paramagnetic (resonance) peak.
The idea of detecting the nematic correlation is clarified in Sec.~\ref{sec:esr_quad},
where we find that various anisotropic interactions result in characteristic angular dependence of $\sin^2\theta\cos^2\theta$
in the quadrupolar TLL phase.
We consider intrachain exchange anisotropies (Sec.~\ref{sec:intrachain_quad}), staggered 
DM interactions (Sec.~\ref{sec:sDM_quad})
and interchain exchange anisotropies (Sec.~\ref{sec:interchain_quad}).
To discuss effects of those anisotropic interactions, we employ the so-called Mori-Kawasaki approach~\cite{MK}.
It requires a reasonable but nontrivial assumption that the paramagnetic peak has a single Lorentzian lineshape.
In fact, in several cases of the standard TLL,
we can justify the assumption based on another approach called a self-energy approach (also known as the Oshikawa-Affleck theory~\cite{OA_PRL,OA_PRB}).
In Sec.~\ref{sec:esr_TLL} we discuss the linewidth of the paramagnetic peak of the standard TLL for two purposes.
One is to compare the angular dependence of the linewidth of the standard TLL with that of the quadrupolar TLL.
The other is to extend the Oshikawa-Affleck theory, originally developed for a single spin chain, to coupled spin-chain systems.
For these purposes,
we first review the Oshikawa-Affleck theory in Sec.~\ref{sec:intrachain} for a longitudinal intrachain anisotropy.
At the same time, we also derive angular dependence of the linewidth based on the Mori-Kawasaki approach
(Sec.~\ref{sec:intrachain_MK}) and see its consistency with the Oshikawa-Affleck theory.
The angular dependence of the linewidth of the standard TLL induced by the staggered DM interaction was derived in Refs.~\onlinecite{OA_PRL, OA_PRB}
and is summarized in Sec.~\ref{sec:sDM}.
Next we show that we can deal with interchain exchange anisotropies using the extended version of the Oshikawa-Affleck theory 
in Sec.~\ref{sec:interchain}.
All these results are briefly given in Table~\ref{table:width}.
Finally, we summarize the paper in Sec.~\ref{sec:summary}.
We also discuss in the Appendix an interesting example of anisotropy to which the self-energy approach is applicable but the Mori-Kawasaki approach is not.

\section{Quadrupolar TLL}\label{sec:quad}

The $S=1/2$ frustrated ferromagnetic chain has the Hamiltonian,
\begin{equation}
 {\mathcal H_{\rm FF}}^0 = \sum_j (J_1\bm S_j \cdot \bm S_{j+1} + J_2 \bm S_j \cdot \bm S_{j+2} - g\mu_BHS_j^z),
  \label{H_FF}
\end{equation}
where $\bm S_j= (S_j^x, S_j^y, S_j^z)$ is an $S=1/2$ spin, 
$J_1<0<J_2$, $g$ and $\mu_B$ are the $g$ factor and the Bohr magneton of electron and $H$ is the magnitude of the magnetic field.
In what follows we take $\hbar$, the Boltzmann constant $k_B$, and the lattice spacing $a$ as unity: $\hbar=k_B=a=1$.
Moreover, we include the factor $g\mu_B$ into $H$ and thus denote $g\mu_BH$ as $H$.

Spin nematic phases emerge usually under a high magnetic field near the saturation field.
If excitation of a bound magnon pair costs lower energy than an unpaired magnon in the fully polarized phase,
reduction of the magnetic field induces a condensation of the bound magnon pair, that is, a quantum phase transition 
from the fully polarized phase to the spin nematic phase.
Let us denote creation and annihilation operators of a bound magnon pair at the $j$th site as $b_j^\dagger$ and $b_j$, respectively.
In the fully polarized phase, creation of the bound magnon pair corresponds to a flipping of neighboring spins represented by an operation 
of $S_j^- S_{j'}^-$,
where $S_j^\pm \equiv S_j^x \pm i S_j^y$.
Thus, a pair flipping operator $S_j^-S_{j'}^-$ corresponds to $b_j^\dag$ and the spin-nematic phase is a condensed phase of these excitations.
In the $S=1/2$ frustrated ferromagnetic chain \eqref{H_FF}, $b_j$ and $b_j^\dag$ are related to the spin operator at $j$th site as follows~\cite{Hikihara_chiral_nematic}.
\begin{align}
 S_j^z &= \frac 12 - 2 b_j^\dagger b_j, 
 \label{Sz2b} \\
 S_j^-S_{j+1}^- &= (-1)^j b_j^\dag.
 \label{Sm2b}
\end{align}

The bound magnon pair is a boson and $b_j$ and $b_j^\dag$ satisfy the canonical commutation relation $[b_j,b_{j'}^\dag]=\delta_{j,j'}$.
In fact, the canonical commutation relation is necessary to respect a commutation relation of spins, 
$[\sum_j S_j^z, \sum_{j'}S_{j'}^+S_{j'+1}^+]=2\sum_jS_j^+S_{j+1}^+$.
We note that the mapping of Eqs.~\eqref{Sz2b} and \eqref{Sm2b} is valid basically in the low-energy limit and that
the bound magnon pair is a hard-core boson 
since $(b_j^\dag)^2=(S_j^-S_{j+1}^-)^2$ vanishes in $S=1/2$ systems.
In general, the collective motion of the hard-core boson in 1D is described by two bosons $\Phi$ and $\Theta$,
which satisfy~\cite{Giamarchi_book, Cazalilla_1dboson_RMP}
\begin{align}
 b_j^\dag &= \biggl(\bar\rho - \frac 1\pi \partial_x\Phi\biggr)^{1/2}  \sum_{n\in\mathbb Z} e^{i2n(\pi\bar\rho x+\Phi)}e^{-i\Theta},
 \label{b2phi} \\
 b_j^\dag b_j &= \biggl(\bar\rho - \frac 1\pi \partial_x\Phi\biggr) \sum_{n=0}^\infty \cos(2\pi n \bar\rho x + 2n\Phi).
 \label{bdb2phi}
\end{align}
$\Phi$ and $\Theta$ are a conjugate of each other related through a commutation relation,
$[\Phi(x), \partial_{x'}\Theta(x')]=i\pi\delta(x-x')$.
$\bar\rho$ is the average density of the bound magnon pair, $\bar\rho = \sum_j \braket{b_j^\dagger b_j}/N$ where $N$ is the number of sites.
Equation~\eqref{Sz2b} relates the magnetization density $M$ and $\bar\rho$,
\begin{equation}
 \bar\rho = \frac 12 \biggl( \frac 12 - M\biggr).
  \label{M2rho}
\end{equation}
It can be rephrased as a relation between the magnetization density and an incommensurate wavenumber of $\braket{S^zS^z}(\omega,q)$
along the spin chain.
Let us denote the wavenumber $q_0$. 
According to Eqs.~\eqref{Sz2b} and \eqref{bdb2phi}, we find that
\begin{equation}
 q_0 = \pi \biggl(\frac 12-M\biggr).
  \label{q02M}
\end{equation}
The validity of the relation \eqref{q02M} is numerically confirmed for quite a wide range of $M$ in the quadrupolar TLL phase~\cite{Onishi_1dnematic}.

\begin{figure}
 \centering
 \includegraphics[bb = 0 0 1000 500, width=\linewidth]{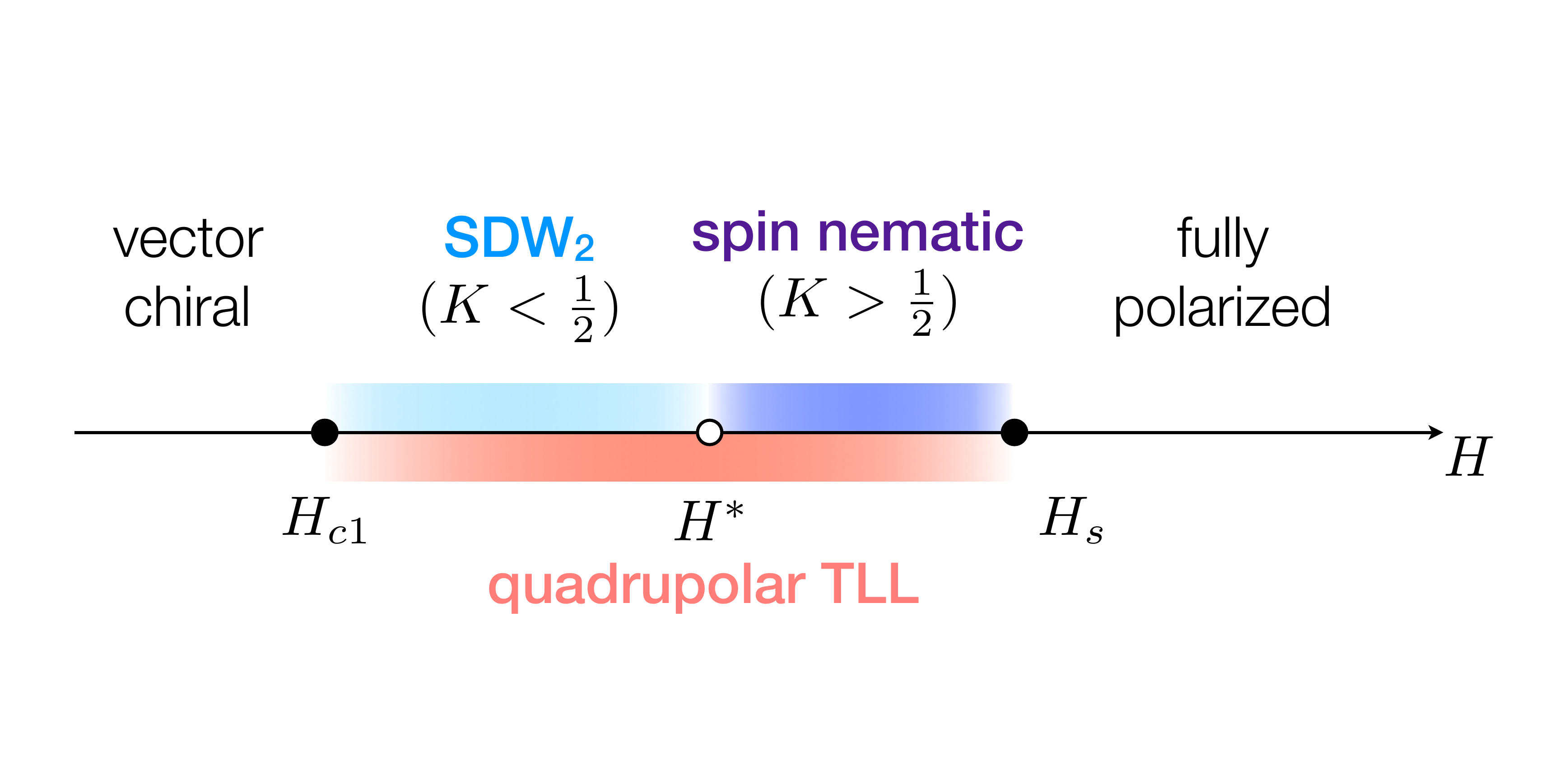}
 \caption{The ground-state phase diagram of the $S=1/2$ frustrated ferromagnetic chain \eqref{H_FF}.
 The quadrupolar TLL phase is spread in a field range $H_{c1}<H<H_s$.
 For $H<H_{c1}$, the ground state belongs to the vector chiral phase.
 For $H>H_s$, the ground state is fully polarized.
 The SDW$_2$ phase and the spin nematic phase are separated at $H=H^\ast$ where $K=1/2$.
 }
 \label{phasediagram_J1J2}
\end{figure}

At low energies,
the $S=1/2$ frustrated ferromagnetic chain is well described by an effective field theory of the quadrupolar 
TLL~\cite{Hikihara_chiral_nematic, Sato_1dnematic_NMR},
\begin{equation}
 {\mathcal H_{\rm FF}}^0 \approx \frac v{2\pi} \int dx \, \biggl(K(\partial_x\Theta)^2 + \frac 1K(\partial_x\Phi)^2\biggr),
  \label{H_qTLL}
\end{equation}
where $K$ is so-called the TLL parameter~\cite{Giamarchi_book}.
In the effective field theory \eqref{H_qTLL}, the unpaired magnon excitation is discarded for a large cost of excitation energy.
The gap of an unpaired magnon is numerically estimated in Refs.~\onlinecite{Sato_quasi1dnematic, Onishi_1dnematic}.
The quadratic TLL has the quasi-long-range nematic order as well as the quasi-long-range spin-density-wave (SDW) order.
This fact can be found in spatial correlations of $\braket{S_r^+S_{r+1}^+S_0^-S_1^-}$ and $\braket{S_r^zS_0^z}$.
They are gradually decaying with a power law of $|r|$~\cite{Hikihara_chiral_nematic}:
\begin{align}
 &\braket{S_r^+S_{r+1}^+S_0^-S_1^-}
 \notag \\
 &= (-1)^r\biggl[\frac{C_0}{|r|^{\frac 1{2K}}} + \frac{C_1\cos(2\pi\bar\rho r)}{|r|^{2K+\frac 1{2K}}}
 +\frac{C_2\cos(4\pi\bar\rho r)}{|r|^{8K+\frac 1{2K}}} + \cdots\biggr],
 \label{SpSp_corr}
\end{align}
and
\begin{align}
 &\braket{S_r^zS_0^z} - M^2 
 \notag \\
 &= -\frac{K}{2\pi^2r^2} + \frac{A_1 \cos(2\pi\bar\rho r)}{|r|^{2K}}
 +\frac{A_2\cos(4\pi\bar\rho r)}{|r|^{8K}}+\cdots,
 \label{Sz_corr}
\end{align}
where $C_n$ and $A_n$ for nonnegative integers $n$ are constants undetermined at the level of the field theory.
$\braket{S_r^xS_0^x}$ and $\braket{S_r^yS_0^y}$ decay exponentially with $|r|$ and
thus the transverse antiferromagnetic order is absent.
The first term of the right hand side of Eq.~\eqref{SpSp_corr} represents the presence of the quasi-long-range nematic order.
Likewise, the second term of the right hand side of Eq.~\eqref{Sz_corr} represents the quasi-long-range SDW order.
The first term proportional to $|r|^{-2}$ merely reflects the fact that the TLL is critical.
When $K>1/2$, the nematic correlation decays slower than the SDW correlation does.
This means that the spin nematic order is more developed than the SDW order.
Thus the spin nematic phase of the $S=1/2$ frustrated ferromagnetic chain is defined as a region of $K>1/2$.
Since the TLL parameter increases monotonically with increase of the magnetization~\cite{Hikihara_chiral_nematic},
the quadrupolar TLL phase is split into two phases: 
an SDW phase ($K<1/2$) on the lower-field side and a spin nematic phase ($K>1/2$) on the higher-field side (Fig.~\ref{phasediagram_J1J2}).
This SDW phase is conventionally referred to as an SDW$_2$ phase.

We emphasize that the SDW$_2$ and the spin nematic phases of the $S=1/2$ frustrated ferromagnetic chain are essentially the same phase,
the quadrupolar TLL phase.
There is no singularity at the boundary between those phases.
In fact, the SDW$_2$ phase has the quasi-long-range spin nematic order and the spin nematic phase has the quasi-long range SDW order.

In Sec.~\ref{sec:esr_quad}, we focus on the SDW$_2$ phase, the low-field region of the quadrupolar TLL phase, because of the following reasons.
First, the SDW$_2$ phase is more easily accessible in experiments including ESR ones.
Second, the qualitative characterization of the quadrupolar TLL (Table~\ref{table:width}) is clearer when the magnetic field is weaker.
We will come back to this point in Sec.~\ref{sec:TH_quad}.

\section{Electron spin resonance}\label{sec:esr}

\subsection{Introduction}

ESR is a unique experimental probe to correlations of electron spins in materials.
It basically probes only uniform correlations at the wavevector $\bm q=0$.
Actually the limitation of the wavevector makes ESR a unique technique sensitive to anisotropy of the spin-spin 
interaction~\cite{OA_PRL, OA_PRB, Furuya_BPCB}.
Therefore, ESR can detect magnetic excitations invisible to other experimental 
techniques~\cite{Zvyagin_CuPM, Umegaki_KCuGaF6, Povarov_ESR_DM, Furuya_boundary, Tiegel_CuPM, Ozerov_ESR_DIMPY}.
Thanks to the sensitivity, the ESR spectroscopy has been used for specifying and modeling anisotropic interactions of electron 
spins~\cite{Furuya_BPCB, Yamada_CuGeO3, Validov_CuPzN}.

ESR experiments measure absorption of microwave going through a target material under the static magnetic field.
According to the linear response theory, 
 absorption intensity $I(\omega)$ is related to a dynamical susceptibility $\chi_{\alpha\alpha}(\omega)$,
\begin{equation}
 I_{\alpha\alpha}(\omega) = \frac{H_R^2\omega}2 \chi''_{\alpha\alpha}(\omega),
  \label{I_w}
\end{equation}
where $H_R$ and $\omega$ are strength and frequency of the oscillating magnetic field transmitting the material.
We denote the direction of the polarization of the oscillating field as $\alpha$ axis.
$\chi''_{\alpha\alpha}(\omega)=-\operatorname{Im}\mathcal G^R_{S^\alpha S^\alpha}(\omega)$ is the imaginary part of the susceptibility and represented in terms of a retarded Green's function of the target material,
\begin{equation}
 \mathcal G^R_{S^\alpha S^\alpha}(\omega) =-\int_0^\infty dt \, e^{i\omega t} \braket{[S^\alpha(t), S^\alpha(0)]}.
\end{equation}
$\bm S=\sum_{\bm r} \bm S_{\bm r}$  is the total spin or the $\bm q=0$ component of the Fourier transform,
$\bm S_{\bm q} = \sum_{\bm r} e^{-i\bm q \cdot \bm r} \bm S_{\bm r}$.
In general, the absorption intensity \eqref{I_w} depends on the polarization of the microwave.

We consider the so-called Faraday configuration where $\hat\alpha$ is perpendicular to the direction of the magnetic field.
Here and in what follows, we denote the unit vector along the $\alpha$ axis as $\hat \alpha$.
If we apply the magnetic field along the $z$ axis, $\hat\alpha$ is obtained from $\hat x$ after a rotation around the $z$ axis.
As far as only the main peak of the ESR spectrum is concerned, which is the case throughout this paper, 
the direction of the polarization within the $xy$ plane is not important [See Eq.~\eqref{polarization}].
Instead of considering $I_{\alpha\alpha}(\omega)$ with $S^\alpha$,
we may deal with a simpler one $I_{+-}(\omega)$ with $S^\pm\equiv S^x\pm i S^y$ as we see below.

Let us consider a system with a Hamiltonian,
\begin{equation}
 \mathcal H = \mathcal H_{\rm SU(2)} -HS^z + \mathcal H',
  \label{H_generic}
\end{equation}
where $\mathcal H_{\rm SU(2)}$ is an SU(2) symmetric (i.e. isotropic) spin-spin interaction,
$-HS^z$ is the Zeeman energy and $\mathcal H'$ is an anisotropic interaction.
All the models we consider in this paper have Hamiltonians of the form \eqref{H_generic}.
Besides, we regard $\mathcal H'$ as a perturbation to the Hamiltonian,
\begin{equation}
 \mathcal H^0 = \mathcal H_{\rm SU(2)} -HS^z.
  \label{H_generic_0}
\end{equation}
The ESR spectrum of the unperturbed system \eqref{H_generic_0} is extremely simple.
Let us denote an unperturbed retarded Green's function as $G^R$.
Likewise, we also use $\mathcal G$ for full Green's functions such as Matsubara and time-ordered ones 
and use $G$ for unperturbed ones throughout the paper.
Interestingly, $G^R_{S^+S^-}(\omega)$ is \emph{exactly} given by
\begin{align}
 G^R_{S^+S^-}(\omega) 
 &= -i\int_0^\infty dt \, e^{i\omega t} \braket{[S^+(t), S^-(0)]}_0
 \label{GR_SpSm} \\
 &= \frac{2\braket{S^z}_0}{\omega-H+i0}.
  \label{GR_SpSm_isotropic}
\end{align}
Here $\braket{\cdot}_0$ means an average with respect to the unperturbed Hamiltonian \eqref{H_generic_0}.
The Green's function \eqref{GR_SpSm_isotropic} immediately leads to
\begin{equation}
  I_{+-}(\omega)=\pi NH_R^2\omega \braket{S^z}_0\delta(\omega-H),
   \label{I_+-_isotropic}
\end{equation}
where $N$ is the number of spins.
$I_{xx}$ and  $I_{yy}$ contain another term proportional to $\delta(\omega+H)$.
However, it is not important because $\omega+H>0$ by definition.
The paramagnetic peak of the unperturbed system is located exactly at $\omega=H$ in the ESR spectrum and
has zero linewidth.

An anisotropic interaction $\mathcal H'$ shifts and broadens the paramagnetic peak~\cite{OA_PRL, Maeda_shift, Furuya_BPCB} 
and even yields an additional absorption peak~\cite{Furuya_boundary, Ozerov_ESR_DIMPY}.
Still, if we focus on the ESR spectrum in the vicinity of $\omega= H$, we can obtain a simple relation,
\begin{equation}
 \mathcal G^R_{S^xS^x}(\omega)\approx \mathcal G^R_{S^yS^y}(\omega) \approx \frac 14\mathcal G^R_{S^+S^-}(\omega).
  \label{polarization}
\end{equation}
A derivation is given in Appendix~\ref{app:polarization}.
The relation \eqref{polarization} is derived from the following identity~\cite{OA_PRB},
\begin{equation}
 \mathcal G^R_{S^+S^-}(\omega) = \frac{2\braket{S^z}}{\omega-H} -\frac{\braket{[\mathcal A, S^-]}}{(\omega-H)^2}
  + \frac 1{(\omega-H)^2}\mathcal G^R_{\mathcal A\mathcal A^\dag}(\omega),
  \label{id}
\end{equation}
where $\omega-H$ is shorthand for $\omega-H+i0$
and $\mathcal A$ is the operator determined from the anisotropic interaction $\mathcal H'$ so that
\begin{equation}
 \mathcal A = [\mathcal H', S^+].
  \label{A_def}
\end{equation}
In the absence of the anisotropy, the identity \eqref{id} immediately reproduces the exact result \eqref{GR_SpSm_isotropic}.
Note that the relation \eqref{polarization} is approximate but the identity \eqref{id} is exact.

\subsection{Mori-Kawasaki approach}

There is a perturbation theory of shift and linewidth of the paramagnetic peak 
called Mori-Kawasaki (MK) theory after an original work of Mori and Kawasaki~\cite{MK}.
In the MK approach, we need to make a single nontrivial assumption that the lineshape of the paramagnetic peak is single Lorentzian.
That is, $\mathcal G^R_{S^+S^-}(\omega)$ is given in the form of
\begin{equation}
 \mathcal G^R_{S^+S^-}(\omega) = \frac{2\braket{S^z}}{\omega - H - \Sigma(\omega)},
  \label{GR_Sigma}
\end{equation}
where $\Sigma(\omega)$ is assumed to be analytic at $\omega=H$.
Given the Green's function \eqref{GR_Sigma},
$\Sigma(\omega)$ is directly related to
the resonance frequency $\omega_r$ and the linewidth $\eta$ of the paramagnetic peak:
$\omega_r = H + \operatorname{Re}\Sigma(H)$ and $\eta = \operatorname{Im}\Sigma(H)$.
One can find a similar argument in a memory function formalism of conductivity~\cite{Giamarchi_memory}.

The assumption of the lineshape is nontrivial although
the lineshape tends to be Lorentzian in systems with a strong exchange interaction at low temperatures~\cite{Anderson_EPR}.
Furthermore, it is quite a subtle problem especially in 1D spin systems whether it has the single Lorentzian 
lineshape~\cite{Dietz_lineshape, ElShawish_KuboTomita}.
The Oshikawa-Affleck theory provided a justification to  the assumption in the $S=1/2$ XXZ spin chain
and the $S=1/2$ Heisenberg spin chain under a staggered magnetic field~\cite{OA_PRL, OA_PRB}.

Once we accept the assumption of the single Lorentzian lineshape, we obtain the following perturbative formulas (MK formulas) for
the resonance frequency $\omega_r$ and the linewidth $\eta$~\cite{OA_PRB},
\begin{align}
 \omega_r - H &\approx - \frac{\braket{[\mathcal A,S^-]}_0}{2\braket{S^z}_0},
 \label{w_r} \\
 \eta &\approx - \frac 1{2\braket{S^z}_0} \operatorname{Im}G^R_{\mathcal A\mathcal A^\dag}(H).
 \label{eta}
\end{align}
Note that we kept leading terms only on the right hand sides of Eqs.~\eqref{w_r} and \eqref{eta}.
Actually we can derive the MK formula \eqref{w_r} for the resonance frequency without relying on
the assumption of the Lorentzian lineshape if the paramagnetic peak is not split~\cite{Maeda_ESR_perturbation}.
In contrast, the validity of the other MK formula \eqref{eta} for the linewidth is less evident.
The validity is confirmed only in limited cases~\cite{OA_PRL, OA_PRB}.
In our case, the assumption is reasonable
because the paramagnetic peak of an $S=1/2$ frustrated ferromagnetic chain compound $\mathrm{LiCuVO_4}$ is well
fitted by the Lorentzian curve at various temperatures~\cite{Validov_LiCuVO4_ESR}
and it is also true for several cases of the standard TLL as we will see later.
On the other hand, we will see in Appendix~\ref{app:uDM} a case where the assumption breaks down.

\subsection{Nematic correlation and ESR}\label{sec:nematic_correlation}

The MK formulas \eqref{w_r} and \eqref{eta} imply that ESR can detect nematic correlations of electron spins.
Those formulas depend crucially on details of $\mathcal A$.
The $\mathcal A$ operator is quadratic in spin operators if $\mathcal H'$ is quadratic.
For the same reason, $[\mathcal A, S^-]$ is also quadratic.

Because spin nematic order parameters are quadratic in spin operators,
Eq.~\eqref{w_r} implies that the resonance frequency is related to a spin nematic order parameter
and Eq.~\eqref{eta} implies that the linewidth is determined from a nematic correlation.
The quadrupolar TLL phase of the $S=1/2$ frustrated ferromagnetic chain has zero spin nematic order parameter
because its spin nematic order is not a long-range but a quasi-long-range one.
For detecting the quasi-long-range spin nematic order of the quadrupolar TLL phase, 
we need to focus on its dynamical aspect.
This is the motivation to consider the ESR linewidth as a probe to the spin nematic order of the $S=1/2$ frustrated ferromagnetic chain.
The implication of the resonance frequency \eqref{w_r} for detecting a long-range spin nematic order parameter
is discussed in detail elsewhere~\cite{furuya_3dnematic}.
Here, let us discuss it only briefly.
An anisotropic exchange interaction on nearest-neighbor bonds, 
$\mathcal H' = \delta \sum_{\langle j, j'\rangle}(S_j^x S_{j'}^x- S_j^y S_{j'}^y)$,
enables us to measure the order parameter of the long-range spin nematic order.
In fact, the anisotropic interaction lead to $[\mathcal A, S^-]= 2\delta \sum_{\langle j,j'\rangle} S_j^- S_{j'}^-$,
which is nothing but the ferroquadrupolar order parameter.

\section{ESR linewidth of the quadrupolar TLL}\label{sec:esr_quad}

Let us take a close look at the linewidth of the $S=1/2$ frustrated ferromagnetic chain with an anisotropic interaction $\mathcal H'$,
\begin{equation}
 \mathcal H_{\rm FF} = {\mathcal H_{\rm FF}}^0 + \mathcal H'.
  \label{H_FF_anisotropic}
\end{equation}
In this section we use the MK formula \eqref{eta}, making the assumption of the single Lorentzian lineshape.
As an anisotropy, we consider an intrachain exchange anisotropy (Sec.~\ref{sec:intrachain_quad}),
a staggered DM interaction (Sec.~\ref{sec:sDM_quad}) and an interchain exchange anisotropy (Sec.~\ref{sec:interchain_quad}).

\subsection{Exchange anisotropy} \label{sec:intrachain_quad}

An exchange anisotropy,
\begin{equation}
 \mathcal H' = \sum_{p=a,b,c}\delta_p \sum_j S_j^pS_{j+1}^p,
  \label{H'_intra}
\end{equation}
is a representative of anisotropic interactions.
Here $(a,b,c)$ denotes the crystalline coordinate.
We call $(x,y,z)$ the laboratory coordinate.
We assume that both of the sets $\{\hat a, \hat b, \hat c\}$
and $\{\hat x, \hat y, \hat z\}$ form right-handed orthogonal coordinate systems.
In what follows we fix the direction of the magnetic field to $\hat z$ and rotate the direction the magnetic field in the crystalline coordinate.

\subsubsection{Angular dependence}\label{sec:theta_quad}

To get insight into the angular dependence of the linewidth, 
we first consider a uniaxial case of  $\delta_a=\delta_b=0$ and next extend it to the general case.
Let $\theta$ be the angle formed by $\hat z$ and $\hat c$.
We may assume that $\hat c$ is on the $zx$ plane.
Then the uniaxial exchange anisotropy is represented in the laboratory coordinate as
\begin{align}
  \mathcal H'
  &= \delta_c \sum_j \bigl[S_j^zS_{j+1}^z \cos^2\theta + S_j^x S_{j+1}^x \sin^2\theta
  \notag \\
  &\quad + (S_j^zS_{j+1}^x+S_j^xS_{j+1}^z)\sin\theta \cos\theta\bigr].
 \label{H'_ex_c}
\end{align}
The operator \eqref{A_def} is thus expressed as
\begin{align}
   \mathcal A
  &= \delta_c \sum_j \bigl[ 
  (S_j^zS_{j+1}^+ + S_j^+S_{j+1}^z)\cos^2\theta 
  \notag \\
  & \quad 
- (S_j^z S_{j+1}^x + S_j^x S_{j+1}^z)\sin^2\theta
  \notag \\
  & \quad + (S_j^+S_{j+1}^+ + S_j^xS_{j+1}^x + S_j^y S_{j+1}^y- 2S_j^zS_{j+1}^z)\sin\theta \cos\theta
  \bigr].
 \label{A_ex_c}
\end{align}
It includes the bound magnon pair annihilation operator $S_j^+S_{j+1}^+\approx (-1)^j b_j$ which leads to a power-law temperature dependence.
The Green's function of  $S_j^zS_{j+1}^z$ also obeys a power law but with a different power.
In contrast, all the other terms such as $S_j^zS_{j+1}^+$ involve creation or annihilation of gapped unpaired magnons.
Green's functions of those operators are exponentially decaying as $e^{-\Delta_1/T}$
and negligible when the temperature is lower than the gap of an unpaired magnon $\Delta_1$,
\begin{equation}
 T < \Delta_1.
  \label{cond_T}
\end{equation}
Therefore, at low temperatures \eqref{cond_T},
we may approximate the operator \eqref{A_ex_c} as
\begin{align}
 \mathcal A
 &\approx \delta_c \sin\theta \cos\theta \sum_j (S_j^+S_{j+1}^+ - 2S_j^zS_{j+1}^z).
 \label{A_ex_c_eff}
\end{align}
Interestingly enough, 
we have already found the angular dependence of the linewidth \eqref{eta}
without calculating details of the correlation function.
Indeed, since $\braket{S^z}_0$ in the numerator of Eq.~\eqref{eta} is independent of $\theta$,
Eq.~\eqref{A_ex_c_eff} gives 
\begin{equation}
 \eta\propto \delta_c^2\sin^2\theta \cos^2\theta.
  \label{width_angle_ex_c}
\end{equation}

\subsubsection{Temperature and field dependences}\label{sec:TH_quad}

We obtained the angular dependence \eqref{width_angle_ex_c} simply by identifying contributions of bound magnon pairs.
In contrast, the temperature and field dependences of the linewidth are more intricate.
Let us look into them under an additional condition,
\begin{equation}
 M \ll 1,
  \label{cond_M}
\end{equation}
for a technical reason.
The condition \eqref{cond_M} is also rephrased as 
\begin{equation}
 \bigl|\bar\rho - \tfrac 14\bigr| \ll 1.
  \label{cond_rho}
\end{equation}
Practically, the condition \eqref{cond_M} can be relaxed to
\begin{equation}
 \max\{M, H/J_2\} \lesssim T/J_2,
  \label{cond_M_relaxed}
\end{equation}
because $\operatorname{Im}G^R_{\mathcal A\mathcal A^\dag}(H)$ is vanishing for $\max\{MJ_2, H\}/T \gg 1$, as we will see later in this section.
Two inequalities \eqref{cond_T} and \eqref{cond_M_relaxed} lead immediately to
\begin{equation}
 \max\{M, H/J_2\} \lesssim \Delta_1/J_2.
  \label{cond_H}
\end{equation}
The field range \eqref{cond_H} turns out to be a low-field region of the quadrupolar TLL phase, which is highly likely to be the SDW$_2$ phase,
because $\Delta_1/H$ does not grow very much with increase of $H$~\cite{Sato_quasi1dnematic, Onishi_1dnematic}.

As the summation of Eq.~\eqref{A_ex_c_eff} indicates,
the linewidth \eqref{eta} picks up
the $\bm q=0$ parts of the correlation functions $\braket{S_j^+S_{j+1}^+S_0^-S_1^-}$ and 
$\braket{S_j^zS_{j+1}^z S_0^z S_1^z}$.
According to the bosonization formulas \eqref{b2phi} and \eqref{bdb2phi},
the $\bm q=0$ part of $S_j^+S_{j+1}^+$ is
\begin{align}
 \sum_j S_j^+S_{j+1}^+
 &\approx \sum_j (-1)^j b_j
 \notag \\
 &\approx \bar\rho \int dx \, e^{i\pi x} \cos(4\pi\bar\rho x + 4\Phi) e^{i\Theta}
 \notag \\
 &=\bar\rho \int dx \, \cos(2\pi Mx-4\Phi)e^{i\Theta},
 \label{SpSp_q=0}
\end{align}
where we picked up the most relevant interaction in the expansion \eqref{b2phi} that compensate the rapid oscillation $(-1)^j=e^{i\pi x}$,
using the assumption \eqref{cond_M}.
The approximation \eqref{SpSp_q=0} breaks down when $M \to 1/2$ because the cosine suffers from the rapid oscillation $e^{i2\pi Mx}\approx e^{i \pi x}$.
According to the bosonization formula \eqref{b2phi},
the correlation function of the operator $\sum_j S_j^+S_{j+1}^+$ has vanishing intensity when $M \to 1/2$ (i.e. $\bar\rho \to 0$).

The $\bm q=0$ part of $S_j^zS_{j+1}^z$ is
\begin{align}
 \sum_j S_j^z S_{j+1}^z
 &\approx \bar\rho^2 \int dx \, \cos(8\pi\bar\rho x + 8\Phi) \notag \\
 &= \bar\rho^2 \int dx \, \cos(4\pi Mx - 8\Phi).
 \label{SzSz_q=0}
\end{align}

The operators $e^{\pm i 4\Phi}e^{i\Theta}$ and $e^{\pm i 8\Phi}$ appearing in Eqs.~\eqref{SpSp_q=0} and \eqref{SzSz_q=0}
are vertex operators with conformal weights $(\Delta_{+}, \bar\Delta_{+})$
and $(\Delta_{z}, \bar\Delta_{z})$, respectively
They are related to the TLL parameter as 
\begin{align}
 (\Delta_{+},\, \bar\Delta_{+}) &= \biggl(\frac{(4K-1)^2}{8K}, \, \frac{(4K+1)^2}{8K}\biggr),
 \label{cweight_+} \\
 (\Delta_{z}, \, \bar\Delta_{z}) &= (8K, \, 8K).
 \label{cweight_z}
\end{align}
Let us suppose that  $\mathcal O(t,x)$ is a vertex operator of the field theory \eqref{H_qTLL} and has a conformal weight $(\Delta, \bar\Delta)$.
Using its retarded Green's function
$G^R_{(\Delta,\bar\Delta)}(\omega,q)=-i\int_0^\infty dt \,e^{i(\omega t-qx)}\braket{[\mathcal O(t,x), \, \mathcal O^\dag(0,0)]}_0$,
we can write $G^R_{\mathcal A\mathcal A^\dag}(\omega)$ for Eq.~\eqref{A_ex_c_eff} as
\begin{align}
 G^R_{\mathcal A\mathcal A^\dag}(\omega)
 &= \frac{N\delta_c^2  \sin^2\theta \cos^2\theta}2
 \notag \\
 &\quad\times \sum_{\sigma=\pm}
 \biggl[ \bar\rho^2 
 G^R_{(\Delta_{+}, \bar\Delta_{+})}(\omega, 2\pi M\sigma)
 \notag \\
 &\quad 
 +4\bar\rho^4
 G^R_{(\Delta_{z}, \bar\Delta_{z})}(\omega, 4\pi M\sigma)
 \biggr].
 \label{GR_AA_ex_c}
\end{align}

The precise form of the retarded Green's function $G^R_{(\Delta,\bar\Delta)}(\omega,q)$ 
is known for general $(\Delta, \bar\Delta)$~\cite{Giamarchi_book}:
\begin{align}
 G^R_{(\Delta,\bar\Delta)}(\omega,q)
 &= -\frac{\sin(2\pi\Delta)}v\biggl(\frac{2\pi T}{v}\biggr)^{2(\Delta+\bar\Delta-1)}
 \notag \\
 & \quad \times B\biggl(\Delta-i\frac{\omega+vq}{4\pi T}, \, 1-2\Delta\biggr)
 \notag \\
 &\quad \times B\biggl(\bar\Delta-i\frac{\omega-vq}{4\pi T}, \, 1-2\bar\Delta\biggr).
\label{GR_vertex_generic}
\end{align}
$B(x,y)=\Gamma(x)\Gamma(y)/\Gamma(x+y)$ is the Beta function and $\Gamma(z)$ is the Gamma function.
Instead of Eq.~\eqref{GR_vertex_generic}, the following equivalent representation is useful for later purpose:
\begin{align}
  & G^R_{(\Delta,\bar\Delta)}(\omega,q)
  \notag \\
  &= -\frac 1{v\sin(2\pi\bar\Delta)\Gamma(2\Delta)\Gamma(2\bar\Delta)} \biggl(\frac{2\pi T}{v}\biggr)^{2(\Delta+\bar\Delta-1)}
  \notag \\
  &\quad\times
  \biggl|\Gamma\biggl(\Delta+i\frac{\omega+vq}{4\pi T}\biggr)\Gamma\biggl(\bar\Delta+i\frac{\omega-vq}{4\pi T}\biggr)\biggr|^2
  \notag \\
  &\quad \times \sin\biggl(\pi \Delta + i\frac{\omega+vq}{4T}\biggr)\sin\biggl(\pi\bar\Delta+i\frac{\omega-vq}{4T}\biggr).
 \label{GR_vertex_generic_2}
\end{align}
We used the identity $\Gamma(z)\Gamma(1-z)=\pi/\sin(\pi z)$ to rewrite it.

The linewidth is determined from $G^R_{\mathcal A\mathcal A^\dag}(H)$ which is governed by 
$G^R_{(\Delta_+,\bar\Delta_+)}(H,\pm 2\pi M)$ and $G^R_{(\Delta_z,\bar\Delta_z)}(H,\pm 4\pi M)$.
Since the Gamma functions in Eq.~\eqref{GR_vertex_generic_2} vanish rapidly for $\max\{|\omega+vq|, |\omega-vq|\}/ T \gg 1$
and the velocity $v$ is of the order of $J_2$~\cite{Sato_quasi1dnematic},
the magnetic field $H$ and the magnetization density $M$ must satisfy the condition \eqref{cond_M_relaxed}.
The condition \eqref{cond_M_relaxed} is directly related to the discussion of the vanishing spatial integral given below Eq.~\eqref{SpSp_q=0}.

Let us ask a question of which of $G^R_{(\Delta_+,\bar\Delta_+)}(H,\pm 2\pi M)$ and 
$G^R_{(\Delta_z,\bar\Delta_z)}(H,\pm 4\pi M)$ governs mostly
the temperature dependence of Eq.~\eqref{GR_AA_ex_c} at $\omega=H$.
The operator $S_j^+S_{j+1}^+$ leads to the power law $(T/v)^{2(\Delta_++\bar\Delta_+)-3}$ 
and the operator $S_j^zS_{j+1}^z$ leads to $(T/v)^{2(\Delta_z+\bar\Delta_z)-3}$.
The latter is negligible compared to the former when $T/v<1$ and $K>1/4\sqrt{3}\approx 0.14$.
The former inequality will be easily satisfied because we limit temperatures to be much lower than the gap of an unpaired magnon
and $\hbar v/a$ is usually larger than the gap.
According to a numerical estimation of $K$~\cite{Hikihara_chiral_nematic}, the inequality for $K$ is also easily satisfied in the SDW$_2$ phase.
Based on this fact, we approximate $G^R_{\mathcal A\mathcal A^\dag}(\omega)$ as
\begin{equation}
 G^R_{\mathcal A\mathcal A^\dag}(\omega) \approx \frac{N\delta_c^2\bar \rho^2\sin^2\theta\cos^2\theta}2 \sum_{\sigma=\pm} G^R_{(\Delta_+,\bar\Delta_+)}(\omega,2\pi M\sigma).
\end{equation}
It immediately follows that
\begin{equation}
 \eta \approx \frac{\delta_c^2\bar\rho^2 \sin^2\theta\cos^2\theta}{2M} \sum_{\sigma=\pm}
  \bigl[-\operatorname{Im}G^R_{(\Delta_+,\bar\Delta_+)}(H,\pm 2\pi M\sigma)\bigr].
  \label{width_ex_c}
\end{equation}
The temperature dependence of the linewidth is determined from those of $G^R_{(\Delta_+,\bar\Delta_+)}(H,\pm 2\pi M)$.
In principle, the temperature dependence of the linewidth tells us the value of $K$ which characterizes the quadrupolar TLL
similarly to the NMR relaxation rate~\cite{Sato_1dnematic_NMR}.
However, it will be challenging to track the intricate temperature dependence of the complicated function \eqref{width_ex_c}
in the narrow 1D phase.
This intricacy motivates us to focus on the angular dependence rather than the temperature dependence.

\subsubsection{General exchange anisotropies}

We have considered the uniaxial anisotropy for simplicity.
Here we extend our discussion to general cases.
Let us rotate the direction of the magnetic field on the $zx$ plane.
For simplicity, we take $\hat b=\hat y$.
Then the $zx$ plane equals to the $ca$ plane and the exchange anisotropy is expressed as
\begin{align}
  \mathcal H'
  &= 
   \sum_j \bigl[
  S_j^zS_{j+1}^z (\delta_c\cos^2\theta + \delta_a\sin^2\theta)
  \notag \\
  &\quad + S_j^xS_{j+1}^x (\delta_c\sin^2\theta +\delta_a\cos^2\theta)
  \notag \\
  & \quad + (S_j^zS_{j+1}^x +S_j^xS_{j+1}^z)(\delta_c-\delta_a)\sin\theta \cos\theta
  \bigr]
  \notag \\
  &\quad +
  \delta_b \sum_j S_j^y S_{j+1}^y.
 \label{H'_ex}
\end{align}
It leads to 
\begin{align}
  \mathcal A
  &= \sum_j \bigl[ (S_j^zS_{j+1}^+ + S_j^+S_{j+1}^z)(\delta_c\cos^2\theta + \delta_a\sin^2\theta)
  \notag \\
  & \quad - (S_j^zS_{j+1}^x + S_j^xS_{j+1}^z)(\delta_c\sin^2\theta +\delta_a\cos^2\theta)
  \notag \\
  &\quad + (S_j^+S_{j+1}^+ +S_j^xS_{j+1}^x+S_j^yS_{j+1}^y - 2S_j^zS_{j+1}^z)
  \notag \\
  &\quad \times (\delta_c-\delta_a)\sin\theta\cos\theta
  \bigr]
 \notag \\
  &\quad -i\delta_b\sum_j (S_j^z S_{j+1}^y + S_j^y S_{j+1}^z).
 \label{A_ex}
\end{align}
Keeping the relevant terms involved only with bound magnon pairs, we find
\begin{equation}
 \mathcal A
  \approx (\delta_c-\delta_a)\sin\theta \cos\theta \sum_j (S_j^+S_{j+1}^+-2S_j^zS_{j+1}^z),
  \label{A_ex_eff}
\end{equation}
and also
\begin{align}
 \eta 
  &\approx \frac{(\delta_c-\delta_a)^2\bar\rho^2 \sin^2\theta\cos^2\theta}{2M} 
 \notag \\
  &\qquad\times \sum_{\sigma=\pm}\bigl[-\operatorname{Im}G^R_{(\Delta_+,\bar\Delta_+)}(H,\pm 2\pi M\sigma)\bigr].
 \label{width_ex}
\end{align}
We found that 
the angular dependence of the linewidth induced by the general (intrachain) exchange anisotropy \eqref{H'_ex} is $\sin^2\theta\cos^2\theta$
and that the anisotropy perpendicular to the plane on which the magnetic field is rotated is negligible.

The $\sin^2\theta\cos^2\theta$ dependence of the linewidth is unique
and seems to be independent of details of anisotropies.
Namely, we may expect that the unique angular dependence of the linewidth characterizes the quadrupolar TLL.
To support this claim, we investigate the linewidth of the quadrupolar TLL induced by other major anisotropies and 
also that of the standard TLL for comparison.
In the rest of this section we deal with the quadrupolar TLL with the staggered DM interaction or with the interchain exchange anisotropy.
The linewidth of the standard TLL is investigated in the next section.

\subsection{Staggered DM interaction}\label{sec:sDM_quad}

The DM interaction is another typical anisotropic interaction,
\begin{equation}
 \mathcal H' = \sum_j \bm D_j \cdot \bm S_j \times \bm S_{j+1}.
  \label{H'_DM}
\end{equation}
When the DM vector alters the direction as $\bm D_j=(-1)^j \bm D$, it is called the staggered DM interaction.
Several spin chain compounds are known to have the staggered DM interaction~\cite{Zvyagin_CuPM, Umegaki_KCuGaF6}.
An $S=1/2$ frustrated ferromagnetic chain compound $\na$ can also have a tiny staggered DM interaction~\cite{Nawa_1dnematic_syn}.

We take $\bm D=D\hat c$.
The staggered DM interaction is actually removable from the Hamiltonian of the $S=1/2$ frustrated ferromagnetic chain \eqref{H_FF_anisotropic}.
A rotation of spin $\bm S_j$ to $\tilde{\bm S}_j$,
\begin{equation}
 \begin{pmatrix}
  S_j^a \\ S_j^b
 \end{pmatrix}
 =
 \begin{pmatrix}
  \cos\alpha & (-1)^j\sin\alpha \\
  -(-1)^j\sin\alpha & \cos\alpha
 \end{pmatrix}
 \begin{pmatrix}
  \tilde S_j^a \\ \tilde S_j^b
 \end{pmatrix},
\end{equation}
eliminates the staggered DM interaction when the angle $\alpha$ equals to $\alpha=D/2|J_1|$~\cite{Affleck_CuBenzoate}.
Instead an exchange anisotropy shows up.
\begin{align}
 \mathcal H_{\rm FF}
 &= \sum_j (J_1\bm S_j \cdot \bm S_{j+1}+ J_2\bm S_j \cdot \bm S_{j+2} -HS_j^z)
 \notag \\
 & \quad + D\sum_j (-1)^j  (S_j^aS_{j+1}^b - S_j^bS_{j+1}^a)
 \notag \\
 &= \sum_j \biggl(-\sqrt{J_1^2+D^2}\tilde{\bm S}_j \cdot \tilde{\bm S}_{j+1} +J_2\tilde{\bm S}_j\cdot \tilde{\bm S}_{j+2} - HS_j^z\biggr)
 \notag \\
 & \quad +\biggl(\sqrt{J_1^2+D^2}-|J_1|\biggr)\sum_j \tilde S_j^c \tilde S_{j+1}^c.
 \label{H_FF_sDM}
\end{align}
The uniaxial anisotropy emerged along the $c$ axis of the crystalline coordinate.
Let us relate the laboratory and the crystalline coordinates.
Here again we consider the rotation $\hat c$ on the $zx$ plane with $\hat b = \hat y$,
\begin{equation}
\left\{
\begin{split}
 \hat x 
 &=-\hat c \sin\theta + \hat a \cos \theta,
 \\
 \hat y
 &= \hat b,
 \\
 \hat z
 &= \hat c \cos\theta + \hat a \sin \theta.
\end{split}
\right.
\end{equation}
Then, up to the first order of $D/|J_1|$, the Hamiltonian \eqref{H_FF_sDM} is approximated as
\begin{align}
  \mathcal H_{\rm FF}
  &\approx \sum_j (J_1\tilde{\bm S}_j\cdot \tilde{\bm S}_{j+1} + J_2\tilde{\bm S}_j \cdot \tilde{\bm S}_{j+2})
  \notag \\
  & \quad - \sum_j (H\tilde S_j^z + (-1)^j h \tilde S_j^y),
\end{align}
with an effective staggered field,
\begin{equation}
 h = \frac{DH}{2|J_1|}\sin\theta.
  \label{h_stag}
\end{equation}
In contrast to the standard TLL phase~\cite{Affleck_CuBenzoate}, 
the staggered field in the quadrupolar TLL phase is irrelevant because it involves unpaired magnons.
Although the staggered field can prevent magnons from forming the quadrupolar TLL by inducing the N\'eel order,
when once the quadrupolar TLL is formed, the weak staggered field has little impact on the Green's function of $G^R_{\mathcal A\mathcal A^\dag}$.

The leading interaction which governs $G^R_{\mathcal A\mathcal A^\dag}$ is the exchange anisotropy, 
\begin{align}
 &\biggl(\sqrt{J_1^2+D^2}-|J_1|\biggr)\sum_j \tilde S_j^c \tilde S_{j+1}^c 
 \notag \\
 &\approx \frac{D^2}{2|J_1|} \sum_j \bigl[
 \tilde S_j^z \tilde S_{j+1}^z\cos^2\theta + \tilde S_j^x \tilde S_{j+1}^x \sin^2\theta
 \notag \\
 & \quad - (\tilde S_j^z \tilde S_{j+1}^x + \tilde S_j^x \tilde S_{j+1}^x)\sin\theta \cos\theta
 \bigr].
\end{align}
The staggered DM interaction as well as the exchange anisotropy induces the linewidth
\eqref{width_ex_c}, where $\delta_c$ is replaced to $D^2/2|J_1|$:
\begin{equation}
 \eta \propto \frac{D^4}{J_1^2} \sin^2\theta \cos^2\theta.
\end{equation}
We have again obtained the dependence of the $\sin^2\theta \cos^2\theta$ type.
Under the additional condition \eqref{cond_rho},
the linewidth is given by
\begin{equation}
  \eta 
  \approx \frac{D^4\bar\rho^2 \sin^2\theta\cos^2\theta}{8J_1^2M} \sum_{\sigma=\pm}[-\operatorname{Im}G^R_{(\Delta_+,\bar\Delta_+)}(H,\pm 2\pi M\sigma)].
 \label{width_sDM}
\end{equation}

\subsection{Interchain interaction}\label{sec:interchain_quad}

Interchain interaction also gives rise to the linewidth~\cite{Furuya_width_interchain}.
The interchain interaction becomes nonnegligible as the temperature is lowered.
Here we show that interchain exchange anisotropies also yield the $\sin^2\theta\cos^2\theta$ dependence of the linewidth.
Note that the effect of the interchain interaction is investigated within the purely 1D phase where the spin chain is independent of the other chains
in the material.

\subsubsection{Unfrustrated interaction}

Including the interchain interaction, we modify our system. 
Here we consider a coupled spin chain system 
where each spin chain is composed of $N_\parallel$ spins and the whole system is composed of $N_\perp=N/N_\parallel$ spin chains.
A Hamiltonian of this system is
\begin{equation}
 \mathcal H = \sum_{\bm R}\mathcal H_{{\rm FF}, \bm R}^0 + \mathcal H_{\rm int},
\end{equation}
where $\mathcal H_{{\rm FF}, \bm R}^0$ and $\mathcal H_{\rm int}$ are the Hamiltonian of the frustrated ferromagnetic chain \eqref{H_FF}
and the interchain interaction, respectively.
The three-dimensional vector $\bm R$ specifies the location of a spin chain.
The spin operator $\bm S_{j,\bm R}$ also acquires the additional index $\bm R$.
Restricting ourselves to the 1D phase, we regard $\mathcal H_{\rm int}$ as a perturbation to the Hamiltonian,
\begin{equation}
 \mathcal H^0 = \sum_{\bm R} \mathcal H_{{\rm FF}, \bm R}^0.
  \label{H_FF_0_interchain}
\end{equation}
The perturbation $\mathcal H_{\rm int}$ contains an isotropic interchain interaction as well as an anisotropic one.
Since the isotropic interaction yields no linewidth, 
we discard it and identify $\mathcal H_{\rm int}$ with an interchain anisotropic interaction $\mathcal H'$.

We consider the most important example of an unfrustrated nearest-neighbor interchain interaction,
\begin{equation}
 \mathcal H' = \sum_{p=a,b,c} \delta_p \sum_j\sum_{\braket{\bm R,\bm R'}} S_{j,\bm R}^p S_{j,\bm R'}^p,
  \label{H'_uf}
\end{equation}
where $\braket{\bm R,\bm R'}$ denotes a combination of nearest-neighbor chains at $\bm R$ and $\bm R'$.
The argument about the nearest-neighbor exchange anisotropy \eqref{H'_uf} given below is generic and it is easy to adapt it to general 
unfrustrated interchain exchange anisotropies.

Let us rotate the direction of the magnetic field parallel to $\hat c$ within the $zx$ plane so that $\hat b=\hat y$:
\begin{align}
  \mathcal H'
  &= 
  \sum_j \sum_{\braket{\bm R,\bm R'}} \bigl[
  S_{j,\bm R}^zS_{j,\bm R'}^z (\delta_c\cos^2\theta + \delta_a\sin^2\theta)
  \notag \\
  &\quad + S_{j,\bm R}^xS_{j,\bm R'}^x (\delta_c\sin^2\theta +\delta_a\cos^2\theta)
  \notag \\
  & \quad + (S_{j,\bm R}^zS_{j,\bm R'}^x +S_{j,\bm R}^xS_{j,\bm R'}^z)(\delta_c-\delta_a)\sin\theta \cos\theta
  \bigr]
  \notag \\
  &\quad +
  \delta_b \sum_j\sum_{\braket{\bm R,\bm R'}} S_{j,\bm R}^y S_{j,\bm R'}^y.
\end{align}
The operator $\mathcal A=[\mathcal H',S^+]$ is 
\begin{align}
  \mathcal A
  &= \sum_j\sum_{\braket{\bm R,\bm R'}} \bigl[ (S_{j,\bm R}^zS_{j,\bm R'}^+ + S_{j,\bm R}^+S_{j,\bm R'}^z)(\delta_c\cos^2\theta + \delta_a\sin^2\theta)
  \notag \\
  & \quad - (S_{j,\bm R}^zS_{j,\bm R'}^x + S_{j,\bm R}^xS_{j,\bm R'}^z)(\delta_c\sin^2\theta +\delta_a\cos^2\theta)
  \notag \\
  &\quad + (S_{j,\bm R}^+S_{j,\bm R'}^+ +S_{j,\bm R}^xS_{j,\bm R'}^x+S_{j,\bm R}^yS_{j,\bm R'}^y - 2S_{j,\bm R}^zS_{j,\bm R'}^z)
  \notag \\
  &\quad \times (\delta_c-\delta_a)\sin\theta\cos\theta
  \bigr]
  \notag \\
  &\quad -i\delta_b\sum_j\sum_{\braket{\bm R,\bm R'}} (S_{j,\bm R}^z S_{j,\bm R'}^y + S_{j,\bm R}^y S_{j,\bm R'}^z).
 \label{A_uf}
\end{align}
The operator $S_{j,\bm R}^+S_{j,\bm R'}^+$ is negligible compared to $S_{j,\bm R}^z S_{j,\bm R'}^z$.
In fact, the nematic correlation function of $S_{j,\bm R}^+S_{j,\bm R'}^+$ for $\bm R\not=\bm R'$ 
is split into a product of two dipolar correlation functions,
\begin{align}
  & \braket{S_{j_1,\bm R_1}^+S_{j_1,\bm R'_1}^+ S_{j_2,\bm R_2}^- S_{j_2,\bm R'_2}^-}_0
  \notag \\
  &= \braket{S_{j_1,\bm R_1}^+S_{j_2,\bm R_1}^-}_0 \braket{S_{j_1,\bm R'_1}^+S_{j_2,\bm R'_1}^-}_0
  \notag \\
  &\quad \times
  (\delta_{\bm R_1,\bm R_2}\delta_{\bm R'_1,\bm R'_2}+\delta_{\bm R_1,\bm R'_2}\delta_{\bm R'_1,\bm R_2}),
\end{align}
which decays exponentially.
Thus we may approximate Eq.~\eqref{A_uf} as
\begin{equation}
 \mathcal A
  \approx -2(\delta_c-\delta_a)\sin\theta \cos\theta\sum_j\sum_{\braket{\bm R,\bm R'}}
  S_{j,\bm R}^z S_{j,\bm R'}^z.
  \label{A_uf_eff}
\end{equation}
The absence of the $S^+S^+$ term is the largest difference from the intrachain interaction \eqref{A_ex_eff}.
The interaction $\sum_j\sum_{\braket{\bm R,\bm R'}}S_{j,\bm R}^zS_{j,\bm R'}^z$ effectively turns into
\begin{align}
 &\sum_{j}\sum_{\braket{\bm R,\bm R'}}S_{j,\bm R}^z S_{j,\bm R'}^z 
 \notag \\
 &\approx \bar\rho^2 \int dx \sum_{\braket{\bm R, \bm R'}} \cos(2\pi\bar\rho x+2\Phi) \cos(2\pi\bar\rho x +2 \Phi')
 \notag \\
 &\approx \frac{\bar\rho^2}2 \int dx \sum_{\braket{\bm R, \bm R'}} \cos\bigl[2(\Phi - \Phi')\bigr],
 \label{A_uf_eff}
\end{align}
where $\Phi$ and $\Phi'$ are the bosonic field of the quadrupolar TLLs at the chain $\bm R$ and at the chain $\bm R'$ respectively
and the rapidly oscillating terms are dropped.

To calculate the retarded Green's function of $e^{i2(\Phi-\Phi')}$,
let us take a brief look at the time-ordered Green's function.
The unperturbed time-ordered correlation function $\braket{T_te^{ia\Phi(t,x)}e^{-ia\Phi(0,0)}}_0$
which we denote as $G^T_{e^{ia\Phi}e^{-ia\Phi}}(t,x)$,
is precisely given by~\cite{Giamarchi_book},
\begin{align}
  &G^T_{e^{ia\Phi}e^{-ia\Phi}}(t,x)
  \notag \\
  &= -\biggl(\frac{\pi T}{2v}\biggr)^{\frac{a^2K}2}
  \biggl[\sinh\biggl(\frac{\pi T }v(x-vt+i\epsilon \operatorname{sgn}(t))\biggr)\biggr]^{-\frac{a^2K}4}
  \notag \\
  &\quad \times   \biggl[\sinh\biggl(\frac{\pi T }v(x+vt+i\epsilon \operatorname{sgn}(t))\biggr)\biggr]^{-\frac{a^2K}4},
 \label{GT_vertex}
\end{align}
where $\epsilon$ is a positive infinitesimal number.
Since the unperturbed Hamiltonian \eqref{H_FF_0_interchain} is free from any interchain interaction, 
the time-ordered Green's function for $e^{i2(\Phi-\Phi')}$ is
simply given by a product $G^T_{e^{i2\Phi}e^{-i2\Phi}}(t,x)G^T_{e^{-i2\Phi'}e^{i2\Phi'}}(t,x)= [G^T_{e^{i2\Phi}e^{-i2\Phi}}(t,x)]^2$.
Equation~\eqref{GT_vertex} tells us that $[G^T_{e^{i2\Phi}e^{-i2\Phi}}(t,x)]^2$ actually equals to
$G^T_{e^{i2\sqrt 2\Phi} e^{-i2\sqrt 2\Phi}}(t,x)$, which is the time-ordered Green's function of 
the vertex operator $e^{i2\sqrt 2\Phi}$ with the conformal weight $(K,K)$.
Considering a fact that a retarded Green's function of a vertex operator is proportional to the imaginary part
of a corresponding time-ordered Green's function~\cite{Giamarchi_book},
we can conclude that the retarded Green's function of $e^{i2(\Phi-\Phi')}$ equals to that of $e^{i2\sqrt 2\Phi}$,
that is, $G^R_{(K,K)}(\omega, q)$.
Thus we obtain
\begin{align}
  &G^R_{\mathcal A\mathcal A^\dag}(\omega)
  \notag \\
  &= \frac{\zeta N\bar\rho^4}2(\delta_c-\delta_a)^2\sin^2\theta \cos^2\theta 
  G^R_{(K, K)}(\omega,  0),
\end{align}
where $\zeta=(\sum_{\braket{\bm R,\bm R'}}1)/N_\perp$ is a half of the number of neighboring spin chains.
The unfrustrated interchain exchange interaction generates the linewidth,
\begin{equation}
  \eta 
  =\frac{\zeta\bar\rho^4(\delta_c-\delta_a)^2\sin^2\theta\cos^2\theta}{4M} \bigl[-\operatorname{Im}G^R_{(K,K)}(H,0)\bigr].
 \label{width_uf_quad}
\end{equation}
The linewidth \eqref{width_uf_quad} also exhibits the angular dependence of $\sin^2\theta\cos^2\theta$.

\subsubsection{Frustrated triangular intearction}

Geometrically frustrated interchain interactions in general affects temperature and angular dependences of the linewidth.
For the standard TLL, we will discuss its effect later in Sec.~\ref{sec:interchain_f}.
The quadrupolar TLL is much simpler.
The frustration has no impact on the temperature and angular dependences of the linewidth.
To see this, we consider a frustrated interchain interaction 
which forms triangular networks of spin chains,
\begin{equation}
 \mathcal H'
  = \sum_{p=a,b,c}\delta_p \sum_j \sum_{\braket{\bm R, \bm R'}} S_{j,\bm R}^p (S_{j,\bm R'}^p + S_{j+1,\bm R'}^p).
  \label{H'_f}
\end{equation}
In the quadrupolar TLL phase,
$\mathcal A=[\mathcal H', S^+]$ of the frustrated interaction \eqref{H'_f} is
\begin{align}
 \mathcal A 
 &\approx -(\delta_c-\delta_a)\sin\theta \cos\theta \sum_j \sum_{\braket{\bm R,\bm R'}}
 \bigl\{S_{j,\bm R}^z (S_{j,\bm R'}^z+S_{j+1,\bm R'}^z) 
 \notag \\
 &\qquad + (\bm R \leftrightarrow \bm R')
 \bigr\} 
 \notag \\
 &\approx -\bar\rho^2(\delta_c-\delta_a)\sin\theta\cos\theta
 \notag \\
 &\quad \times \int dx \, \sum_{\braket{\bm R,\bm R'}} \bigl\{1+\cos(2\pi\bar\rho)\bigr\}
 \cos\bigl[2(\Phi-\Phi')\bigr] .
\end{align}
Thus the frustrated interchain interaction yields the linewidth of
\begin{align}
  \eta
  &=\frac{\zeta\bar\rho^4(\delta_c-\delta_a)^2\sin^2\theta\cos^2\theta}{2M}  \bigl\{1+\cos(2\pi\bar\rho)\bigr\}^2
  \notag \\
  &\quad \times \bigl[-\operatorname{Im}G^R_{(K,K)}(H,0)\bigr].
\end{align}
which is identical with the linewidth \eqref{width_uf_quad} induced by the unfrustrated interchain interaction except for the minor modification of
the coefficient.

\subsection{Short summary and discussion}

We have dealt with the linewidth of the quadrupolar TLL induced by three major anisotropic interactions:
the exchange anisotropies within a chain [Eq.~\eqref{width_ex}]
and between chains [Eq.~\eqref{width_uf_quad}] and the staggered DM interaction [Eq.~\eqref{width_sDM}].
The linewidth of the paramagnetic peak of the quadrupolar TLL induced by these anisotropic interactions exhibits the angular dependence
of $\sin^2\theta \cos^2\theta$ (Table~\ref{table:width}).

The angular dependence of $\sin^2\theta\cos^2\theta$ follows from the simple fact that 
the $\bm q=0$ component of correlation functions involved with unpaired magnons
are negligible compared to those of bound magnon pairs.
This argument is independent of dimensionality and applicable to spin nematic phases in higher-dimensional systems.
Moreover, in higher-dimensional systems, 
the paramagnetic resonance frequency \eqref{w_r} will give an order parameter of long-range spin nematic order~\cite{furuya_3dnematic}.

We did not deal with an important case of the uniform DM interaction because of the following reason.
Our argument in this section relies crucially on the assumption that the paramagnetic peak has the \emph{single} Lorentzian lineshape.
As we discuss in Appendix~\ref{app:uDM}, this assumption breaks down in the standard TLL of a system with a uniform DM interaction.
It is unclear whether the uniform DM interaction also splits the paramagnetic peak of the quadrupolar TLL.
We keep as an open problem investigation of effects of the uniform DM interaction on ESR of the quadrupolar TLL.

\section{ESR of the standard TLL}\label{sec:esr_TLL}

This section has a twofold aim.
First, we investigate the ESR linewidth of the standard TLL for comparison with that of the quadrupolar TLL.
Second, we extend the Oshikawa-Affleck theory and discuss the linewidth of coupled TLL systems induced by interchain exchange anisotropies.
In Ref.~\onlinecite{Furuya_width_interchain}, the author used the MK formula \eqref{eta}
to investigate the linewidth of the standard TLL induced by interchain exchange
anisotropies without justifying the assumption of the lineshape.
Using the extended Oshikawa-Affleck theory,
we prove that its lineshape affected by interchain interactions is indeed the single Lorentzian one.

\subsection{Non-Abelian bosonization}\label{sec:nonabelian}

In the previous sections we dealt with the $S=1/2$ frustrated ferromagnetic chain \eqref{H_FF_anisotropic} and discussed the quadrupolar TLL with the aid of
the Abelian bosonization technique \eqref{b2phi} and \eqref{bdb2phi}.
Here, in order to discuss the standard TLL, we investigate an $S=1/2$ Heisenberg antiferromagnetic (HAFM) chain
described by a Hamiltonian,
\begin{equation}
 \mathcal H_{\rm HAFM} = {\mathcal H_{\rm HAFM}}^0 + \mathcal H',
  \label{H_HAFM_anisotropic}
\end{equation}
with
\begin{equation}
 {\mathcal H_{\rm HAFM}}^0 =  J \sum_j \bm S_j \cdot \bm S_{j+1} - HS^z,
  \label{H_HAFM_0}
\end{equation}
and an anisotropic perturbation $\mathcal H'$ to it.
Moreover, we use a non-Abelian bosonization technique to describe the standard TLL instead of the Abelian one.
Although those bosonization techniques are equivalent, the non-Abelian one is practically more convenient in this Section.

Let us start our discussion from the fully SU(2) symmetric case, that is, the unperturbed system \eqref{H_HAFM_0} at zero magnetic field.
Its low-energy effective Hamiltonian is given by
\begin{equation}
 \mathcal H^0 = \frac v2 \int dx\, \bigl\{(\partial_x\tilde\phi)^2 + (\partial_x\phi)^2\bigr\}.
  \label{H_TLL_0}
\end{equation}
Here $v$ is the velocity of the TLL and
$\phi$ and $\tilde\phi$ are U(1) compactified boson fields of the TLL.
They are compactified as
\begin{equation}
 \left\{
  \begin{split}
   \phi &\sim \phi + 2\pi NR, \\
   \tilde\phi &\sim \tilde\phi + \frac{\tilde N}R,
  \end{split}
 \right.
 \label{compact}
\end{equation}
with a compactification radius $R=1/\sqrt{2\pi}$ and $N,\tilde N\in\mathbb Z$.
The symbol $\sim$ means an identification through the compactification.
The Green's function of $\phi$ is related to that of $\tilde\phi$ thanks to the following duality of $\phi$ and $\tilde\phi$.
Since they satisfy a commutation relation $[\phi(x), \partial_{x'}\tilde\phi'(x)]=i\delta(x-x')$,
they are subject to equations of motion given by
\begin{equation}
 \partial_0 \phi = \partial_1\tilde\phi, \quad \partial_0 \tilde\phi = -\partial_1\phi,
  \label{eom_dual}
\end{equation}
with $\partial_0 = v^{-1}\partial_t$ and $\partial_1 =\partial_x$.
The two fields $\phi$ and $\tilde\phi$ are dual in the sense of Eq.~\eqref{eom_dual}.

The effective Hamiltonian \eqref{H_TLL_0} results from the bosonization of the spin chain.
The non-Abelian bosonization formulas are given by
\begin{align}
 S_j^z &= \frac 1{\sqrt{8\pi^2}}(J_R^z + J_L^z) +C_s \cos(\pi x) n^z,
 \label{Sz2J} \\
 S_j^\pm &= \frac 1{\sqrt{8\pi^2}} (J_R^\pm + J_L^\pm) + C_se^{i\pi x} n^\pm.
 \label{Spm2J}
\end{align}
Here $\bm J$ is the SU(2) current and $\bm J_R$ and $\bm J_L$ are its right-moving and left-moving components
and $\bm n$ is a field corresponding to the N\'eel order $(-1)^j\bm S_j$.
The constant $C_s$ is nonuniversal and thus undetermined within the field theory.
The chiral SU(2) currents $\bm J_R$ and $\bm J_L$ have conformal weights $(\Delta,\bar\Delta)=(0,1)$ and $(1,0)$ respectively and
$\bm n$ has the weight $(\Delta,\bar\Delta)=(\frac 14, \frac 14)$.
Note that the sum $\Delta+\bar\Delta$ is the scaling dimension of the operator that determines its relevance in the RG sense.

In terms of the SU(2) current, we can rewrite the Hamiltonian \eqref{H_TLL_0} so that the SU(2) symmetry is more explicit:
\begin{equation}
 \mathcal H^0 
 = \frac v{48\pi} \int dx \, (\bm J_R \cdot \bm J_R + \bm J_L \cdot \bm J_L).
  \label{H_TLL_0_su2}
\end{equation}
Translation rules from the SU(2) fields $\bm J$ and $\bm n$ to the U(1) fields $\phi$ and $\tilde\phi$ are the followings.
\begin{equation}
\left\{
\begin{split}
 J_R^z &= \sqrt{4\pi}\,(-\partial_0 + \partial_1) \varphi_R,
 \\
 J_L^z &= \sqrt{4\pi}\,(\partial_0 + \partial_1) \varphi_L,
 \\
 J_R^\pm &= \sqrt 2 \, e^{\pm i\sqrt{8\pi}\varphi_R},
 \\
 J_L^\pm &= \sqrt 2 \, e^{\mp i\sqrt{8\pi}\varphi_L}.
\end{split}
\right.
\label{J2phi}
\end{equation}
and
\begin{equation}
 \left\{
  \begin{split}
   n^x &= \cos\sqrt{2\pi}\tilde\phi, \\
   n^y &= \sin\sqrt{2\pi}\tilde\phi, \\
   n^z &= \cos\sqrt{2\pi}\phi.
  \end{split}
 \right.
 \label{n2phi}
\end{equation}
The chiral fields $\varphi_R$ and $\varphi_L$ represent the right-moving and left-moving parts of the TLL,
\begin{equation}
 \left\{
\begin{split}
 \phi(t,x) &= \varphi_R(x-vt) + \varphi_L(x+vt), \\
 \tilde\phi(t,x) &= \varphi_R(x-vt) - \varphi_L(x+vt).
\end{split}
\right.
\end{equation}
We discarded in the Hamiltonian \eqref{H_TLL_0_su2} a marginally irrelevant interaction in the RG sense, $g\bm J_R \cdot \bm J_L$ with $g\propto J$.
Since this interaction is isotropic, we may discard it as far as we focus on ESR induced by
perturbations $\mathcal H'$ which keep the chiral symmetry.
We will investigate in Appendix~\ref{app:uDM} a case where the condition of the chirality breaks down.

The Zeeman energy $-HS^z$ turns effectively into
\begin{equation}
 -HS^z = - \frac H{\sqrt{8\pi^2}} \int dx \, (J_R^z+J_L^z) = - \frac H{\sqrt{2\pi}}\int dx \, \partial_x\phi.
\end{equation}
The Zeeman energy is absorbed into the quadratic Hamiltonian \eqref{H_TLL_0} by shifting
\begin{equation}
 \left\{
  \begin{split}
   \varphi_R &\to \varphi_R + \frac 1{\sqrt{8\pi}} \frac{Hx}v,
   \\
   \varphi_L &\to \varphi_L + \frac 1{\sqrt{8\pi}} \frac{Hx}v.
  \end{split}
 \right.
 \label{shift_phi}
\end{equation}
The shift \eqref{shift_phi} modifies the non-Abelian bosonization formulas \eqref{Sz2J} and \eqref{Spm2J} into
\begin{align}
 S_j^z
 &= \frac H{2\pi v} + \frac 1{\sqrt{8\pi^2}} (J_R^z + J_L^z)
 \notag \\
 &\quad + C_s \bigl\{ \cos[(\pi +H/v)x] n^z - \sin[(\pi + H/v)x]\e \bigr\},
 \label{Sz2J_H} \\
 S_j^\pm
 &= \frac 1{\sqrt{8\pi^2}} (J_R^\pm e^{\pm iHx/v} + J_L^\pm e^{\mp i Hx/v})
 +C_s e^{i\pi x} n^\pm,
 \label{Spm2J_H}
\end{align}
where $\e=\sin\sqrt{2\pi}\phi$ corresponds to the dimerization $(-1)^j \bm S_j\cdot \bm S_{j+1}$.
In particular, $S^\pm$ is represented as
\begin{equation}
 S^\pm = \frac 1{\sqrt{8\pi^2}} \int dx \, (J_R^\pm e^{\pm iHx/v} + J_L^\pm e^{\mp iHx/v}).
  \label{Spm_J}
\end{equation}
In addition, the Zeeman energy affects the compactification radius $R$ in Eq.~\eqref{compact} because it reduces the SU(2) symmetry to 
U(1)~\cite{Bogolioubov_K, Affleck_CuBenzoate}.
Nevertheless, we may neglect this effect in the ESR problem when the magnetic field is weak~\cite{OA_PRB}.
The Zeeman energy thus results only in the shift of $\phi$.
As a result, even in the presence of the magnetic field,
correlation functions of $\braket{J_\mu^aJ_\nu^b}$ for $\mu,\nu=R,L$ satisfies simple SU(2) symmetric relations,
\begin{equation}
 \braket{J_\mu^a J_\nu^b}_0 = \delta_{\mu,\nu}\delta^{a,b} \braket{J_\mu^z J_\mu^z}_0.
  \label{JJ_su2}
\end{equation}
The same relation holds for the retarded Green's function of the SU(2) currents.

The relation \eqref{Spm_J} allows us to write the unperturbed retarded Green's function $G^R_{S^+S^-}(\omega)$ in terms of 
$G^R_{(\Delta,\bar\Delta)}(\omega, q)$ [Eq.~\eqref{GR_vertex_generic}].
\begin{align}
 G^R_{S^+S^-}(\omega) 
 &= \frac N{4\pi^2} \biggl[
 G^R_{(0,1)}(\omega, -H/v) + 
  G^R_{(1,0)}(\omega, H/v)
 \biggr] \notag \\
 &= \frac N{2\pi^2} G^R_{(1,0)}(\omega,H/v) \notag \\
 &\approx \frac{NH}{\pi v} \frac 1{\omega-H+i0}.
 \label{GR_SpSm_isotropic_TLL}
\end{align}
We approximated $\omega\approx H$ in the last line.
According to Eq.~\eqref{Sz2J_H}, the total magnetization is given by  $\braket{S^z}_0=NH/2\pi v$.
Substituting it into Eq.~\eqref{GR_SpSm_isotropic_TLL}, we reproduce the exact result \eqref{GR_SpSm_isotropic}.

\subsection{Oshikawa-Affleck theory: longitudinal exchange anisotropy}\label{sec:intrachain}

Now that we confirmed that the non-Abelian bosonization approach reproduces the exact result \eqref{GR_SpSm_isotropic} for the 
unperturbed system \eqref{H_HAFM_0}, we move on to investigation of influence of anisotropic interactions $\mathcal H'$.
In what follows we discuss the linewidth in two independent ways: the self-energy approach and the MK approach.
The self-energy approach is also known as the Oshikawa-Affleck theory.
The self-energy approach is important for providing a justification to the assumption of the lineshape
even though its application scope is more limited than the MK approach.

For later convenience we review the Oshikawa-Affleck theory for the linewidth caused by the 
uniaxial intrachain exchange anisotropy,
\begin{equation}
 \mathcal H' = \delta \sum_j S_j^z S_{j+1}^z.
\end{equation}
In the language of the non-Abelian bosonization \emph{at zero magnetic field}, it becomes effectively 
\begin{equation}
 \mathcal H' = \lambda \int dx \, J_R^z J_L^z,
  \label{H'_l_TLL}
\end{equation}
with $\lambda\propto \delta$.
The magnetic field shifts $J_R^z$ and $J_L^z$ by an amount of $H/\sqrt{2}$ [Eq.~\eqref{shift_phi}] and
modifies $\mathcal H'$ into
\begin{align}
 \mathcal H'
 &= \lambda \int dx \, J_R^z J_L^z + \frac{\lambda H}{\sqrt 2} \int dx \, (J_R^z + J_L^z).
\end{align}
The second term is an effective magnetic field generated by the anisotropic interaction and
causes a shift of the resonance frequency from $\omega=H$ to $\omega = H(1-2\pi\lambda)$.
Thus it has no impact on the linewidth.
Below we discuss the influence of the quadratic interaction \eqref{H'_l_TLL} on the linewidth.

For a while we set $v=1$ and recover it in the final result [Eq.~\eqref{width_l}] from dimensional analysis.
The correlation function $\braket{S^+S^-}(\omega)$ is written in terms of the SU(2) current as
\begin{align}
  \braket{S^+S^-}(\omega) 
  &= \frac N{8\pi^2} \bigl[ \braket{J_R^x J_R^x}(\omega,-H) + \braket{J_L^x J_L^x}(\omega,H)
  \notag \\
  &\quad + \braket{J_R^y J_R^y}(\omega,-H) + \braket{J_L^y J_L^y}(\omega,H)
  \bigr].
 \label{SpSm_JJ}
\end{align}
Oshikawa and Affleck used a trick to develop the self-energy approach.
Performing a $\pi/2$ rotation on the $az$ plane,
they rewrote the correlation $\braket{J_R^aJ_R^a}$ as $\braket{J_R^zJ_R^z}$ in the rotated system.
The correlation of the rotated system is denoted as $\braket{J_R^zJ_R^z}_{a\to z}$.
The rotation simplifies $\braket{S^+S^-}$ to
\begin{align}
 &\braket{S^+S^-}(\omega) \notag \\
 &= \frac N{8\pi^2} \bigl[ \braket{J_R^zJ_R^z}_{x \to z}(\omega,-H) + \braket{J_L^zJ_L^z}_{x\to z}(\omega,H)
 \notag \\
 &\quad + \braket{J_R^zJ_R^z}_{y\to z}(\omega,-H) + \braket{J_L^zJ_L^z}_{y\to z}(\omega,H)
 \bigr]
 \notag \\
 &= \frac{N(\omega + H)^2}{4\pi} \braket{\phi\phi}_{x\to z}(\omega,H)
 \notag  \\
 &\quad + \frac{N(\omega+H)^2}{4\pi} \braket{\phi\phi}_{y\to z}(\omega,H).
 \label{SpSm_JJ_rot}
\end{align}
Now our task is reduced to calculation of $\braket{\phi\phi}$ in the presence of rotated perturbations.
Thus the retarded Green's function $\mathcal G^R_{S^+S^-}(\omega)$ is derived from that of $\phi$,
\begin{align}
 \mathcal G^R_{\phi\phi}(\omega,q) 
 &= -i\int_0^\infty dt \, \int_{-\infty}^\infty dx \, e^{i(\omega t - qx)} 
 \braket{[\phi(t,x), \phi(0,0)]} \notag \\
 &= \frac 1{\omega^2 - q^2 - \Pi^R(\omega,q)}.
\end{align}
$\Pi^R(\omega,q)$ is the self-energy of the retarded Green's function of $\phi$.
Writing the self-energy of $\mathcal G^R_{\phi\phi}$ in the rotated system as $\Pi^R_{a\to z}$, 
we can write the Green's function $\mathcal G^R_{S^+S^-}$ as follows.
\begin{equation}
  \mathcal G^R_{S^+S^-}(\omega)
  = \frac{N(\omega+H)^2}{4\pi}\sum_{a=x,y} \frac 1{\omega^2 - H^2 - \Pi^R_{a\to z}(\omega,H)}.
\end{equation}
Near $\omega=H$, it is approximated as
\begin{equation}
  \mathcal G^R_{S^+S^-}(\omega)
  \approx \frac{NH}{2\pi } \frac 1{\omega - H - \frac 1{2H}\Pi^R_{a\to z}(H,H)}.
  \label{GR_SpSm_Lorentzian}
\end{equation}
The self-energy $\Pi^R_{a\to z}$ is related to the linewidth of the paramagnetic peak.
If $\Pi^R_{x\to z} = \Pi^R_{y\to z}$, the paramagnetic resonance peak is composed of the single Lorentzian peak with the linewidth,
\begin{equation}
 \eta = -\frac 1{2H} \operatorname{Im}\Pi^R_{a\to z}(H,H).
  \label{eta_Pi}
\end{equation}

The imaginary-time formalism is more convenient in derivation of the self-energy.
The retarded Green's function $\mathcal G^R_{\phi\phi}(\omega,q)$ is obtained from
a corresponding Matsubara Green's function,
\begin{equation}
 \mathcal G_{\phi\phi}(i\omega_n,q) = \frac{1}{(i\omega_n)^2-q^2-\Pi(i\omega_n,q)},
\end{equation}
after analytic continuation $i\omega_n \to \omega+i0$.
The self-energy $\Pi^R(\omega,q)$ is also obtained via the analytic continuation:
$\Pi^R(\omega,q) = \Pi(i\omega_n\to\omega+i0, q)$.

The self-energy $\Pi(i\omega_n,q)$ is obtained as follows.
Let $\bm r$ be a coordinate $\bm r=(\tau,x)$ of the Euclidean spacetime.
The Matsubara Green's function ${\mathcal G_{\phi\phi}}^{a\to z}(\bm r)$ under the rotation $a\to z$ 
satisfies the Dyson equation,
\begin{align}
 &{\mathcal G_{\phi\phi}}^{a\to z}(\bm r)
 \notag \\
 &=  G_{\phi\phi}(\bm r) 
 \notag \\
 & \quad
 +\int d\bm r_1 d\bm r_2 \, G_{\phi\phi}(\bm r_1) \Pi_{a\to z}(\bm r_2-\bm r_1) {\mathcal G_{\phi\phi}}^{a\to z}(\bm r-\bm r_3) 
 \notag \\
 &=  G_{\phi\phi}(\bm r) 
 \notag \\
 & \quad
 +\int d\bm r_1 d\bm r_2\, G_{\phi\phi}(\bm r_1) \Pi_{a\to z}(\bm r_2-\bm r_1) G_{\phi\phi}(\bm r-\bm r_3) 
 \notag \\
 & \quad + \cdots.
 \label{Dyson}
\end{align}
$\Pi_{a\to z}(\bm r)$ is the inverse Fourier transform of $\Pi_{a\to z}(i\omega_n, q)$.

In the present case, we obtain a perturbative expression of $\Pi_{x\to z}(\bm r)$ as follows.
The rotation $x\to z$ transforms the longitudinal anisotropy \eqref{H'_l_TLL} to 
\begin{align}
 {\mathcal H'}_{x\to z} 
 &= \lambda \int dx \, J_R^x J_L^x \notag \\
 &= \lambda \int dx \, \bigl(\cos\sqrt{8\pi}\phi + \cos\sqrt{8\pi}\tilde\phi\bigr).
 \label{H'_x2z_l}
\end{align}
Up to the second order of $\lambda$,
two cosines $\cos\sqrt{8\pi}\phi$ and $\cos\sqrt{8\pi}\tilde\phi$ can be dealt with independently to 
calculate the self-energy $\Pi_{x\to z}(\bm r)$ because of $\braket{e^{ia\phi}e^{ib\tilde\phi}}_0=0$ for any $a,b\in\mathbb R$~\cite{Giamarchi_book}.
Moreover, $\cos\sqrt{8\pi}\phi$ and $\cos\sqrt{8\pi}\tilde\phi$ give exactly the same contribution to $\Pi_{x\to z}$
thanks to the duality \eqref{eom_dual}~\cite{OA_PRB}.

The perturbation \eqref{H'_x2z_l} has no contribution to the self-energy at the first order of $\lambda$ 
because of $\braket{e^{\pm i\sqrt{8\pi}\phi}}_0=\braket{e^{\pm i\sqrt{8\pi}\tilde\phi}}_0=0$.
At the second order of $\lambda$,
the cosine $\lambda\cos\sqrt{8\pi}\phi$ enters into the expansion \eqref{Dyson} as follows.
\begin{align}
 & \int d\bm r_1\bm dr_2 \, G_{\phi\phi}(\bm r_1) \Pi_{x\to z}(\bm r_2-\bm r_1)G_{\phi\phi}(\bm r-\bm r_2)
 \notag \\
 &=\frac{\lambda^2}{2!}
  \int d\bm r_1 d\bm r_2 \,\sum_{\sigma=\pm}\frac 14 
 \braket{\phi(\bm 0) e^{i\sigma\sqrt{8\pi}\phi(\bm r_1)} e^{-i\sigma\sqrt{8\pi}\phi(\bm r_2)} \phi(\bm r)}_0 \notag \\
 &=\pi\lambda^2 \sum_{\sigma=\pm} \int d\bm r_1d\bm r_2 \, 
 \notag \\
 & \quad\times
  \bigl[
 \braket{\phi(\bm 0)\phi(\bm r_1)}_0
 \braket{e^{i\sigma\sqrt{8\pi}\phi(\bm r_1)}e^{-i\sigma\sqrt{8\pi}\phi(\bm r_2)}}_0
 \braket{\phi(\bm r_2)\phi(\bm r)}_0 
 \notag \\
 & \quad 
+ \braket{\phi(\bm 0)\phi(\bm r_2)}_0 \braket{e^{i\sigma\sqrt{8\pi}\phi(\bm r_1)}e^{-i\sigma \sqrt{8\pi}\phi(\bm r_2)}}_0
 \braket{\phi(\bm r_1)\phi(\bm r)}_0
 \notag \\
 & \quad - \braket{\phi(\bm 0)\phi(\bm r_1)}_0 \braket{e^{i\sigma\sqrt{8\pi}\phi(\bm r_1)}e^{-i\sigma \sqrt{8\pi}\phi(\bm r_2)}}_0
 \braket{\phi(\bm r_1)\phi(\bm r)}_0
 \notag \\
 & \quad - \braket{\phi(\bm 0)\phi(\bm r_2)}_0 \braket{e^{i\sigma\sqrt{8\pi}\phi(\bm r_1)}e^{-i\sigma \sqrt{8\pi}\phi(\bm r_2)}}_0
 \braket{\phi(\bm r_2)\phi(\bm r)}_0
 \bigr].
 \label{Dyson_2nd}
\end{align} 
Here we used the Wick's theorem.
Performing the Fourier transform on the both hand sides,
we extract the following self-energy
\begin{equation}
 \Pi_{x\to z}(i\omega_n, q) = 4\pi\lambda^2 \bigl[G_{(1,1)}(i\omega_n,q) - G_{(1,1)}(0,0)\bigr].
\end{equation}
Including the contribution from $\cos\sqrt{8\pi}\tilde\phi$,
we obtain
\begin{equation}
 \Pi_{x\to z}(i\omega_n, q) = 8\pi\lambda^2 \bigl[G_{(1,1)}(i\omega_n,q) - G_{(1,1)}(0,0)\bigr].
\end{equation}
and
\begin{equation}
 \Pi^R_{x\to z}(\omega,q) = 8\pi\lambda^2 \bigl[G^R_{(1,1)}(\omega,q) - G^R_{(1,1)}(0,0)\bigr].
\end{equation}
The imaginary part of the retarded Green's function in a limit, $\max\{|\omega-q|, |\omega+q|\}\ll T$, is easily obtained
from Eq.~\eqref{GR_vertex_generic_2}.
Keeping the leading term only, we find
\begin{equation}
 \operatorname{Im}G^R_{(1,1)}(\omega,q) \approx - \pi^2\omega T.
\end{equation}
and
\begin{equation}
 \Pi^R_{x\to z}(H,H) =8\pi^3 \lambda^2 HT.
\end{equation}

It is obvious from these calculations that ${\mathcal H'}_{y\to z} = \lambda \int dx \, (-\cos\sqrt{8\pi}\phi +\cos\sqrt{8\pi}\tilde\phi)$
gives rise to the identical self-energy $\Pi^R_{y\to z}(H,H)=8\pi^3\lambda^2HT$.
In the end, we found that the paramagnetic peak is the single Lorentzian peak with the linewidth
\begin{equation}
 \eta = \frac{4\pi^3\lambda^2 T}{v^2},
  \label{width_l}
\end{equation}
derived initially by Oshikawa and Affleck~\cite{OA_PRB}.
Here we recovered $v$ from the dimensional analysis.
The formulation reviewed here is straightforwardly extended to quasi-one-dimensional systems in Sec.~\ref{sec:interchain}.

\subsection{Mori-Kawasaki approach: general angles}\label{sec:intrachain_MK}

Here we employ the MK approach to calculate the linewidth induced by an exchange anisotropy
\begin{align}
 \mathcal H' 
 &= \delta \sum_j S_j^c S_{j+1}^c
 \\
 &= \delta \sum_j \bigl[ S_j^z S_{j+1}^z \cos^2\theta + S_j^x S_{j+1}^x \sin^2\theta
 \notag \\
 & \quad - (S_j^zS_{j+1}^x + S_j^x S_{j+1}^z) \sin\theta \cos \theta
 \bigr],
 \label{H'_ex_general}
\end{align}
and check its consistency with the Oshikawa-Affleck theory \eqref{width_l}.
Equation~\eqref{H'_ex_general} represents a uniaxial exchange anisotropy along the $c$ axis.
In terms of the bosonized effective field theory, $\mathcal H'$ is expressed as
\begin{widetext}
\begin{align}
  \mathcal H'
  &= \lambda \cos^2\theta \int dx \,  J_R^z J_L^z
  +\frac{\lambda\sin^2\theta}4 \int dx \, (J_R^+J_L^+ + J_R^- J_L^- + J_R^+J_L^- e^{i2Hx/v} + J_R^- J_L^+ e^{-i2Hx/v})
  \notag \\
  &\quad 
  +\frac{\lambda\sin\theta\cos\theta}2 \int dx \, \bigl\{ (J_R^zJ_L^- + J_R^+ J_L^z)e^{iHx/v}
  + (J_R^z J_L^+ + J_R^- J_L^z)e^{-iHx/v} \bigr\}.
\end{align}
Here we discarded the linear terms with respect to $\bm J_R$ or $\bm J_L$ that do not contribute to the linewidth.

To obtain $\mathcal A$ defined as the equal-time commutator \eqref{Spm_J}, 
we need to know equal-time commutation relations of the SU(2) currents.
The equal-time commutation relations immediately follow from their operator product expansions (Appendix~\ref{app:ope}):
\begin{equation}
 \frac 1{\sqrt{8\pi^2}} [J_{R/L}^a(x), J_{R/L}^b(y)] = \frac i{\sqrt{2\pi}} \delta^{ab}\partial_y\delta(x-y) + if^{abc} J_{R/L}^c(y)\delta(x-y),
  \label{comm_JJ}
\end{equation}
with the completely antisymmetric tensor $f^{abc}$ with $f^{xyz}=1$.
The first term of Eq.~\eqref{comm_JJ} represents the chiral anomaly.
Effects of the chiral anomaly seem not to be discussed explicitly in Ref.~\onlinecite{OA_PRB}.
However, at the end of the day, it turns out to be negligible.
Here we simply ignore the chiral anomaly and discuss it in Appendix~\ref{app:anomaly}.

Then the operator $\mathcal A$ is 
\begin{align}
  \mathcal A
  &= \lambda\biggl(\cos^2\theta-\frac{\sin^2\theta}2\biggr) \int dx \, (J_R^+J_L^z e^{iHx/v} + J_R^zJ_L^+ e^{-iHx/v})
  -\frac{\lambda\sin^2\theta}2 \int dx \, (J_R^z J_L^- e^{iHx/v} + J_R^-J_L^z e^{-iHx/v})
  \notag \\
  &\quad + \frac{\lambda\sin\theta\cos\theta}2 \int dx \, (2J_R^+J_L^+ -4 J_R^z J_L^z + J_R^+ J_L^- e^{i2Hx/v} + J_R^-J_L^+ e^{-i2Hx/v}).
 \label{A_JJ}
\end{align}
All the operators of the form $J_R^aJ_L^b$ for $a=x,y,z$ are an operator with a conformal weight $(1,1)$.
Counting the number of the operator, we find
\begin{align}
  G^R_{\mathcal A\mathcal A^\dag}(\omega)
  &= N\lambda^2 \biggl[2\biggl(\cos^2\theta - \frac{\sin^2\theta}2\biggr)^2
  \bigl\{G^R_{(1,1)}(\omega, -H/v) + G^R_{(1,1)}(\omega,H/v) \bigr\}
  \notag \\
  &\qquad 
  +2\biggl(\frac{\sin^2\theta}2\biggr)^2 
    \bigl\{G^R_{(1,1)}(\omega, -H/v) + G^R_{(1,1)}(\omega,H/v) \bigr\}
  \notag \\
  & \qquad + \biggl(\frac{\sin\theta\cos\theta}2\biggr)^2 \bigl\{
  32 G^R_{(1,1)}(\omega,0) + 4 G^R_{(1,1)}(\omega,2H/v) + 4G^R_{(1,1)}(\omega,-2H/v)\bigr\}
  \biggr].
 \label{GR_AA_ex_TLL}
\end{align} 
When $H/T\ll 1$, the imaginary part of $G^R_{(1,1)}(H,q)$ is independent of $q$.
According to Eq.~\eqref{GR_vertex_generic_2}, we may approximate it up to the leading order of $H/T$ as
\begin{equation}
 \operatorname{Im}G^R_{(1,1)}(H,q) = - \frac{\pi^2HT}{v^3}.
\end{equation}
The insensitivity to the wavenumber simplifies the imaginary part of Eq.~\eqref{GR_AA_ex_TLL}.
\begin{align}
 \operatorname{Im} G^R_{\mathcal A\mathcal A^\dag}(\omega=H)
 &=N\lambda^2 [-\operatorname{Im}G^R_{(1,1)}(H,0)] 
 \biggl\{ 4\biggl(\cos^2\theta - \frac{\sin^2\theta}2\biggr)^2
 +4\biggl(\frac{\sin^2\theta}2\biggr)^2
 + 40\biggl(\frac{\sin\theta\cos\theta}2\biggr)^2
 \biggr\} \notag \\
 &= -\frac{2N\lambda^2 \pi^2HT}{v^3} (1+\cos^2\theta).
\end{align}
\end{widetext}
It leads to
\begin{equation}
 \eta = \frac{2\pi^3 \lambda^2 T}{v^2} (1+\cos^2\theta).
  \label{width_general}
\end{equation}
For $\theta=0$, the result \eqref{width_general} is identical to the linewidth \eqref{width_l} by the longitudinal exchange anisotropy.

Thus far we did not include a renormalization effect of the conformal weight by the magnetic field.
In other words, we did not take into account a fact that
$\braket{S^+_jS^-_0}_0$ decays more slowly than $\braket{S^z_jS^z_0}_0$ under strong magnetic fields.
The renormalization effect affects the angular dependence of the linewidth \eqref{width_general}.
However, both correlation functions are algebraically decaying different from the quadrupolar TLL.
The qualitative feature of the angular dependence \eqref{width_general} will be kept unchanged 
up to a finite magnetic field.
The upper limit of the magnetic field is given similarly to the case \eqref{cond_M_relaxed} of the quadrupolar TLL.
That is, since a Gamma function of the Green's function \eqref{GR_vertex_generic_2} of a vertex operator 
gives rise to a sizable magnitude of the linewidth only for $\max\{|\omega+vq|, |\omega-vq|\}/T \lesssim 1$,
the magnetic field $H$ and the magnetization density $M$ needs to satisfy
\begin{equation}
 \max\{M, H/J\} \lesssim T/J,
  \label{cond_M_relaxed_TLL}
\end{equation}
to make the linewidth finite.

\subsection{Staggered DM interaction}\label{sec:sDM}

The angular dependence of the linewidth of the $S=1/2$ HAFM chain with the staggered DM interaction 
\begin{equation}
 \mathcal H' = \sum_j (-1)^j\bm D \cdot \bm S_j \times \bm S_{j+1}.
\end{equation}
was already discussed in Refs.~\onlinecite{OA_PRL, OA_PRB},
\begin{equation}
 \eta = \frac 1{16} \sqrt{\frac \pi 2} \biggl(\frac{\Gamma(\frac 14)}{\Gamma(\frac 34)}\biggr) \frac{Jh^2}{T^2} \ln (J/T),
  \label{width_stag}
\end{equation}
where $h$ is a staggered magnetic field effectively generated from the staggered DM and given by
\begin{equation}
 h = \frac{DH\sin\theta}{2J}.
\end{equation}
It leads to the angular dependence,
\begin{equation}
 \eta \propto \sin^2\theta.
\end{equation}
The staggered DM interaction as well as the exchange anisotropy leads to the angular dependence with periodicity $\pi$.

\subsection{Unfrustrated interchain interaction: self-energy approach}\label{sec:interchain}

Recently the author discussed the linewidth induced by interchain exchange anisotropies using the MK approach
assuming the single Lorentzian lineshape~\cite{Furuya_width_interchain}.
Here we confirm that the assumption is true for a simple case.
Let us consider a system with the following Hamiltonian,
\begin{equation}
  \mathcal H_{\rm Q1D}
  = J \sum_{j, \bm R} \bm S_{j,\bm R}\cdot \bm S_{j+1,\bm R} -HS^z 
  +  \mathcal H'.
 \label{H_q1d}
\end{equation}
The last term $\mathcal H'$ represents interchain interactions.
Since isotropic interaction does not contribute to the linewidth,
we regard $\mathcal H'$ as anisotropic interchain exchange interactions,
for example, a transverse unfrustrated one,
\begin{equation}
 \mathcal H' = \delta_\perp \sum_{\braket{\bm R,\bm R'}}\sum_j S_{j,\bm R}^xS_{j,\bm R'}^x.
  \label{H'_uf_x}
\end{equation}
The unperturbed Hamiltonian is 
\begin{equation}
 {\mathcal H_{\rm Q1D}}^0 = J\sum_{j,\bm R} \bm S_{j,\bm R}\cdot \bm S_{j+1,\bm R} - HS^z.
  \label{H_q1d_0}
\end{equation}
At low energies compared to $J$, it effectively turns into
\begin{equation}
 {\mathcal H_{\rm Q1D}}^0 = \sum_{\bm R}\frac v2 \int dx \, \{(\partial_x\tilde\phi_{\bm R})^2
  + (\partial_x\phi_{\bm R})^2\}.
  \label{H_q1d_TLL_0}
\end{equation}
The boson field $\phi_{\bm R}$ and its dual $\tilde\phi_{\bm R}$ describe the TLL on a spin chain specified by $\bm R$.
At low energies, the perturbation \eqref{H'_uf_x} becomes
\begin{align}
 \mathcal H'
 &= \lambda_\perp \sum_{\braket{\bm R,\bm R'}}\int dx \, n_{\bm R}^x n_{\bm R'}^x,
 \label{H'_uf_x_TLL}
\end{align}
with $\lambda_\perp\propto\delta_\perp$.
Different from the intrachain exchange anisotropy,
the staggered component of the spin gives the leading term of the perturbation \eqref{H'_uf_x_TLL}.

To discuss the lineshape,
we need to extend the Oshikawa-Affleck theory to the case of $N_\perp$ independent spin chains perturbed by interchain interactions.
Now the total $S^+$ is
\begin{equation}
 S^+ = \sum_{\bm R} \int dx \, (J_{R,\bm R}^+ e^{iHx/v} + J_{L,\bm R}^+ e^{-iHx/v}).
\end{equation}
In what follows we set $v=1$ again for a while.
The Fourier transform $\braket{S^+S^-}(\omega)$ of the correlation $\braket{S^+(t)S^-(0)}$ is
reduced to that of a single boson $\Phi_0=\sum_{\bm R}\phi_{\bm R}/\sqrt{N_\perp}$,
\begin{align}
 &\braket{S^+S^-}(\omega)
 \notag \\
 &= \frac{N_\parallel}{8\pi^2} \sum_{\bm R,\bm R'} \bigl[ \braket{J_{R,\bm R}^x J_{R,\bm R'}^x}(\omega,-H)
 + \braket{J_{L,\bm R}^x J_{L,\bm R'}^x}(\omega,H) 
 \notag \\
 &\quad
 +\braket{J_{R,\bm R}^y J_{R,\bm R'}^y}(\omega,-H)
 + \braket{J_{L,\bm R}^y J_{L,\bm R'}^y}(\omega,H) 
 \bigr]
 \notag \\
 &= \frac{N_\parallel(\omega+H)^2}{4\pi} \sum_{\bm R,\bm R'} \bigl[ \braket{\phi_{\bm R}\phi_{\bm R'}}_{x\to z}(\omega,H)
 \notag \\
 & \quad 
 + \braket{\phi_{\bm R}\phi_{\bm R'}}_{y\to z}(\omega, H)
 \bigr]
 \notag \\
 &= \frac{N(\omega+H)^2}{4\pi} \bigl[ \braket{\Phi_0 \Phi_0}_{x\to z}(\omega,H)
 + \braket{\Phi_0 \Phi_0}_{y\to z}(\omega,H)
 \bigr].
\end{align}
To write the Hamiltonian in terms of $\Phi_0$,
we consider a recombination of $\{\phi_{\bm R}\}_{\bm R}$.
For simplicity of notation, we rename them to be $\phi_0, \phi_1, \phi_2,\cdots, \phi_{N_\perp-1}$.
A recombination 
\begin{align}
 \Phi_0 &= \frac{\phi_0+\phi_1+ \cdots + \phi_{N_\perp-1}}{\sqrt{N_\perp}} , \\
 \Phi_1 &= \frac{\phi_0 -\phi_1}{\sqrt 2}, \\
 \Phi_2 &= \frac{\phi_0 + \phi_1 - 2\phi_2}{\sqrt 6}, \\
 & \vdots \notag \\
 \Phi_{N_\perp-1} &= \frac{\phi_0+ \phi_1 + \cdots + \phi_{N_\perp-2} - (N_\perp-1)\phi_{N_\perp-1}}{\sqrt{N_\perp(N_\perp-1)}},
\end{align}
keeps the unperturbed Hamiltonian \eqref{H_q1d_TLL_0} invariant:
\begin{align}
 {\mathcal H_{\rm Q1D}}^0 = \sum_{m=0}^{N_\perp-1} \frac 12 \int dx \, \{
 (\partial_x\tilde\Phi_m)^2 + (\partial_x\Phi_m)^2 \}.
 \label{H_q1d_TLL_0_rot}
\end{align}
Since $(\Phi_0,\tilde\Phi_0)$ is decoupled from any other $(\Phi_{m\not=0},\tilde\Phi_{m\not=0})$,
 we are able to write the Green's function $\mathcal G^R_{S^+S^-}(\omega)$
in terms of $\mathcal G^R_{\Phi_0\Phi_0}(\omega,q)$ in the same fashion as the single-chain case,
\begin{equation}
 \mathcal G^R_{S^+S^-}(\omega) = \frac{N H}{2\pi}\sum_{a=x,y} \frac 1{\omega-H- \frac 1{2H}\Pi^R_{a\to z}(\omega,H)}.
\end{equation}

Let us derive the self-energy $\Pi^R_{x\to z}(\omega,q) = \Pi_{x\to z}(i\omega_n \to \omega +i0,q)$.
The perturbation \eqref{H'_uf_x_TLL} is composed of two cosines.
\begin{align}
 {\mathcal H'}_{x\to z}
 &= \lambda_\perp \sum_{\braket{\bm R,\bm R'}} \int dx \, \cos\sqrt{2\pi}\phi_{\bm R}\cos\sqrt{2\pi}\phi_{\bm R'}
 \notag \\
 &= \frac{\lambda_\perp}2 \sum_{\braket{\bm R,\bm R'}}  \int dx \, \bigl\{
 \cos\sqrt{4\pi}\Phi_+ + \cos\sqrt{4\pi}\Phi_-
 \bigr\},
 \label{H'_uf_x2z}
\end{align}
with $\Phi_\pm= (\phi_{\bm R} \pm \phi_{\bm R'})/\sqrt 2$.
The second-order term of the expansion \eqref{Dyson} is given by
\begin{widetext}
\begin{align}
 &\int d\bm r_1d\bm r_2 G_{\Phi_0\Phi_0}(\bm r_1) \Pi_{x\to z}(\bm r_2-\bm r_1) G_{\Phi_0\Phi_0}(\bm r-\bm r_3) \notag \\
 &=\frac 1{2!}\biggl(\frac{\lambda_\perp}2\biggr)^2\int d\bm r_1 d\bm r_2 \sum_{\braket{\bm R,\bm R'}} \sum_{\sigma=\pm}  \sum_{\sigma' = \pm}
 \frac 14 \braket{\Phi_0(\bm 0) e^{i\sigma \sqrt{4\pi}\Phi_{\sigma'}(\bm r_1)} e^{-i\sigma\sqrt{4\pi} \Phi_{\sigma'}(\bm r_2)}
  \Phi_0(\bm r)}_0
 \notag \\
 &=\frac{\pi\lambda^2}8 \int d\bm r_1d\bm r_2 \sum_{\braket{\bm R,\bm R'}}\sum_{\sigma=\pm} \sum_{\sigma'=\pm}\,
  \bigl[
 \braket{\Phi_0(\bm 0)\Phi_{\sigma'}(\bm r_1)}_0
 \braket{e^{i\sigma\sqrt{4\pi}\Phi_{\sigma'}(\bm r_1)}e^{-i\sigma\sqrt{4\pi}\Phi_{\sigma'}(\bm r_2)}}_0
 \braket{\Phi_{\sigma'}(\bm r_2)\Phi_0(\bm r)}_0 
 \notag \\
 & \quad 
+ \braket{\Phi_0(\bm 0)\Phi_{\sigma'}(\bm r_2)}_0 \braket{e^{i\sigma\sqrt{4\pi}\Phi_{\sigma'}(\bm r_1)}e^{-i\sigma \sqrt{4\pi}\Phi_{\sigma'}(\bm r_2)}}_0
 \braket{\Phi_{\sigma'}(\bm r_1)\Phi_0(\bm r)}_0
 \notag \\
 & \quad - \braket{\Phi_0(\bm 0)\Phi_{\sigma'}(\bm r_1)}_0 \braket{e^{i\sigma\sqrt{4\pi}\Phi_{\sigma'}(\bm r_1)}e^{-i\sigma \sqrt{4\pi}\Phi_{\sigma'}(\bm r_2)}}_0
 \braket{\Phi_{\sigma'}(\bm r_1)\Phi_0(\bm r)}_0
 \notag \\
 & \quad - \braket{\Phi_0(\bm 0)\Phi_{\sigma'}(\bm r_2)}_0 \braket{e^{i\sigma\sqrt{4\pi}\Phi_{\sigma'}(\bm r_1)}e^{-i\sigma \sqrt{4\pi}\Phi_{\sigma'}(\bm r_2)}}_0
 \braket{\Phi_{\sigma'}(\bm r_2)\Phi_0(\bm r)}_0
 \bigr].
 \label{Dyson_2nd_uf}
\end{align} 
\end{widetext}
The unperturbed Hamiltonian \eqref{H_q1d_TLL_0} is invariant under exchange of an arbitrary pair of $\phi_{\bm R}$ and $\phi_{\bm R'}$.
This symmetry leads to $\braket{\Phi_0 \Phi_-}_0=0$ and $\braket{\Phi_0 \Phi_+}_0 = \sqrt{2/N_\perp}\braket{\Phi_0\Phi_0}_0$.
Using these relations, we obtain the self-energy,
\begin{equation}
 \Pi^R_{x\to z}(\omega,q)
  = \pi\lambda^2 \zeta \bigl[F(\omega,q) - F(0,0)\bigr],
  \label{Pi_uf}
\end{equation}
where $\zeta$ is the number of nearest-neighbor spin chains per chain.
$F(\omega,q)$ is the Fourier transform of the retarded Green's function 
$G^R_{e^{i\sqrt{4\pi}\Phi_{+}}e^{-i\sqrt{4\pi}\Phi_{+}}}(t,x)$, 
which equals to 
$[G^R_{e^{i\sqrt{2\pi}\phi_{\bm R}}e^{-i\sqrt{2\pi}\phi_{\bm R}}}(\omega,q)]^2=G^R_{e^{i\sqrt{4\pi}\phi_{\bm R}}e^{-i\sqrt{4\pi}\phi_{\bm R}}}(\omega,q)=G^R_{(\frac 12, \frac 12)}(\omega,q)$ according to the same argument below Eq.~\eqref{A_uf_eff}.
Furthermore, the self-energy $\Pi^R_{y\to z}(\omega,q)$ is exactly identical to Eq.~\eqref{Pi_uf}.
It follows from the duality of $\phi_{\bm R}$ and $\tilde\phi_{\bm R}$ and from
\begin{equation}
 {\mathcal H'}_{y\to z} = \frac{\lambda_\perp}2 \sum_{\braket{\bm R,\bm R'}} \int dx \, \bigl\{ \cos\sqrt{4\pi}\tilde\Phi_+
  + \cos\sqrt{4\pi}\tilde\Phi_-\bigr\},
\end{equation}
with $\tilde\Phi_\pm=(\tilde\phi_{\bm R}\pm \tilde\phi_{\bm R'})/\sqrt 2$.
Therefore, substituting $\operatorname{Im}G^R_{(\frac 12, \frac 12)}(H,H)\approx-\pi^2H/4T$ [Eq.~\eqref{GR_vertex_generic_2}] for $H/T\ll 1$
into the self-energy $\Pi^R_{x\to z}(H,H)$, we obtain
\begin{equation}
 \eta = \frac{\pi^3\zeta \lambda^2}{8T}.
  \label{width_uf_l}
\end{equation}
The self-energy approach turned out to work similarly to the case of the intrachain exchange anisotropy.
It gives a consistent result with the MK approach~\cite{Furuya_width_interchain} and
also a justification of the assumption of the Lorentzian lineshape.

\subsection{Unfrustrated interchain interaction: MK approach}\label{sec:interchain_MK}

Now we apply the MK formula \eqref{eta} to an interchain version of the anisotropy \eqref{H'_ex_general}, that is,
\begin{align}
  \mathcal H'
  &= \delta_\perp \sum_{\braket{\bm R,\bm R'}} \sum_j \bigl[ S_{j,\bm R}^z S_{j,\bm R'}^z \cos^2\theta
  + S_{j,\bm R}^x S_{j, \bm R'}^x \sin^2\theta 
  \notag \\
  &\quad + (S_{j,\bm R}^z S_{j,\bm R'}^x+S_{j,\bm R}^x S_{j,\bm R'}^x)\sin\theta\cos\theta
  \bigr],
  \label{H'_uf_standard}
\end{align}
or in the non-Abelian bosonization language,
\begin{widetext}
\begin{align}
  \mathcal H'
  &=\lambda_\perp \sum_{\braket{\bm R,\bm R'}} \int dx \, \bigl[ \bigl\{
  n_{\bm R}^z \cos(Hx/v) - \e_{\bm R}\sin(Hx/v)  \bigr\}\bigl\{
  n_{\bm R'}^z\cos(Hx/v) - \e_{\bm R'}\sin(Hx/v)
  \bigr\}\cos^2\theta
  +n_{\bm R}^x n_{\bm R'}^x \sin^2\theta
  \notag \\
  &\quad +\bigl[ \bigl\{n_{\bm R}^z \cos(Hx/v) - \e_{\bm R}\sin(Hx/v)\bigr\} n_{\bm R'}^x 
  +n_{\bm R}^x \bigl\{n_{\bm R'}^z \cos(Hx/v) - \e_{\bm R'}\sin(Hx/v)\bigr\}
  \bigr]\sin\theta \cos\theta
  \bigr].
  \label{H'_uf_TLL}
\end{align} 
The next task is to take a commutator with $S^+$.
Equal-time commutators involved with $\bm n$ and $\e$ follow from operator product expansions (Appendix~\ref{app:ope}),
\begin{equation}
 \left\{
 \begin{split}
  \frac 1{\sqrt{8\pi^2}} [n^x(x), J^+_{R/L}(y)] 
 &=  -\frac 12 \bigl\{n^z(x)\pm i \e(x)\bigr\}\delta(x-y), \\ 
 \frac 1{\sqrt{8\pi^2}} [n^z(x), J^+_{R/L}(y)] 
 &= \frac 12 n^+(x)\delta(x-y), \\
 \frac 1{\sqrt{8\pi^2}} [\e(x), J^+_{R/L}(y)] 
 &= \pm \frac i2 n^+(x)\delta(x-y),  
 \end{split}
       \right.
\end{equation}
where $R$ and $L$ on the left hand side correspond to the upper and lower signs on the right hand side, respectively.
These commutation relations lead to the equal-time commutator $\mathcal A$,
\begin{align}
   \mathcal A
   &= \lambda_\perp \sum_{\braket{\bm R, \bm R'}} \biggl[ (\cos^2\theta - \sin^2\theta)
   \bigl\{(n_1^x n_2^z + n_1^z n_2^x)\cos(Hx/v) - (n_1^x \e_2 + \e_1 n_2^x)\sin(Hx/v)\bigr\}
   \notag \\
   &\quad +i\cos^2\theta \bigl\{
   (n_1^y n_2^z + n_1^z n_2^y) \cos(Hx/v) - (n_1^y \e_2 + \e_1 n_2^y)\sin(Hx/v)
   \bigr\}
   \notag \\
   &\quad +\sin\theta \cos\theta \bigl\{ 2n_1^x n_2^x + i(n_1^x n_2^y + n_1^y n_2^x)
   -n_1^z n_2^z - \e_1 \e_2
   \notag \\
   &\quad -(n_1^zn_2^z-\e_1\e_2)\cos(2Hx/v) - (n_1^z\e_2+\e_1n_2^z)\sin(2Hx/v)
   \bigr\}
   \biggr]
\end{align}

The unperturbed Hamiltonian \eqref{H_q1d_TLL_0} has the ${\rm SU(2) \times SU(2)\simeq SO(4)}$ symmetry, 
a combination of the SU(2) symmetry of the right mover and that of the left mover.
The SO(4) symmetry manifests itself in the following relation.
Let us introduce a four-dimensional vector $\bm N = (N^0, N^1, N^2, N^3) = (\e, n^x, n^y, n^z)$.
Then a retarded Green's function $G^R_{N^a_{\bm R}, N^b_{\bm R'}}(t,x)$ satisfies the SO(4) symmetric relation,
\begin{equation}
 G^R_{N^a_{\bm R}, N^b_{\bm R'}}(t,x) = \delta_{\bm R,\bm R'} \delta^{a,b} G^R_{N^3_{\bm R}, N^3_{\bm R}}(t,x)
  = \frac 12 \delta_{\bm R,\bm R'} \delta^{a,b} G^R_{(\frac 14, \frac 14)}(t,x).
\end{equation}
The following relation is also important.
\begin{equation}
 G^R_{N^a_{\bm R}N^b_{\bm R'}, N^a_{\bm R}N^b_{\bm R'}}(t,x)
  = \biggl(\frac 12\biggr)^2 G^R_{(\frac 12, \frac 12)}(t,x),
\end{equation}
for $\bm R\not=\bm R'$.
The SO(4) symmetry and the decoupling of spin chains in the unperturbed system simplifies calculations of $G^R_{\mathcal A\mathcal A^\dag}(\omega)$.
Counting the number of operators simply leads to
\begin{align}
 G^R_{\mathcal A\mathcal A^\dag}(\omega)
 &=N \zeta \lambda_\perp^2 \biggl[ \frac{ (\cos^2\theta - \sin^2\theta)^2 + \cos^4\theta}2
 \bigl\{G^R_{(\frac 12, \frac 12)}(\omega,H/v) + G^R_{(\frac 12, \frac 12)}(\omega,-H/v)\bigr\}
 \notag \\
 &\quad + \frac{\sin^2\theta \cos^2\theta}2 \bigl\{4G^R_{(\frac 12, \frac 12)}(\omega, 0)
 + G^R_{(\frac 12, \frac 12)}(\omega,2H/v)+ G^R_{(\frac 12, \frac 12)}(\omega, -2H/v)\bigr\}
 \biggr] .
\end{align}
\end{widetext}
It follows from the imaginary part $\operatorname{Im}G^R_{(\frac 12, \frac 12)}(H,q) \approx - \pi^2H/4Tv$ for $H/T\ll 1$ that
\begin{equation}
 \operatorname{Im}G^R_{\mathcal A\mathcal A^\dag}(H) = \frac{\pi^2H N\zeta\lambda_\perp^2}{8T} \bigl(2\cos^4\theta + \sin^2\theta\bigr).
\end{equation}
Combining it with $\braket{S^z}_0= NH/2\pi v$, we obtain
\begin{equation}
 \eta = \frac{\pi^3 \zeta \lambda_\perp^2}{8T}\bigl(2\cos^4\theta + \sin^2\theta).
  \label{width_uf}
\end{equation}
The result \eqref{width_uf} is identical to the linewidth \eqref{width_general} obtained by the self-energy approach for $\theta=\pi/2$.
The linewidth \eqref{width_uf} becomes maximum at $\theta=0 \mod \pi$ as well as the intrachain one.
From the period of the angle that maximizes the linewidth, we can distinguish the quadrupolar TLL and the standard TLL.

\subsection{Frustrated triangular interchain interaction}\label{sec:interchain_f}

Frustrated interchain interactions result in a different angular dependence.
To see this, we consider the following interchain interaction,
\begin{widetext}
\begin{align}
  \mathcal H'
  &= \frac{\delta_\perp}2 \sum_{\braket{\bm R, \bm R'}} \sum_j \bigl[ 
  S_{j,\bm R}^z (S_{j,\bm R'}^z + S_{j+1,\bm R'}^z)\cos^2\theta
  + S_{j,\bm R}^x (S_{j,\bm R'}^x + S_{j+1,\bm R'}^x)\sin^2\theta 
 \notag \\
  &\quad + \bigl\{ S_{j,\bm R}^z (S_{j,\bm R'}^x +S_{j+1,\bm R'}^x) 
 + S_{j,\bm R}^x (S_{j,\bm R'}^z + S_{j+1,\bm R'}^z)
  \bigr\}\sin\theta \cos\theta 
 \notag \\
  &\quad + (\bm R \leftrightarrow \bm R')
  \bigr].
 \label{H'_f_standard}
\end{align}
At low energies the interaction \eqref{H'_f} turns into
\begin{align}
  \mathcal H'
  &= -\frac{\lambda_\perp H}{2v} \sum_{\braket{\bm R,\bm R'}}\int dx \, \bigl[
  \bigl\{(n^z_{\bm R}n^z_{\bm R'}-\e_{\bm R}\e_{\bm R'})\sin(2Hx/v)
  +(n^z_{\bm R}\e_{\bm R'}+\e_{\bm R}n^z_{\bm R'})\cos(2Hx/v)
  \bigr\}\cos^2\theta 
 \notag \\
  &\quad + \bigl\{ (n^x_{\bm R}n^z_{\bm R'}+n^z_{\bm R}n^x_{\bm R'})\sin(Hx/v)
  +(n^x_{\bm R}\e_{\bm R'} + \e_{\bm R}n^x_{\bm R})\cos(Hx/v)
  \bigr\}\sin\theta\cos\theta
  \bigr].
 \label{H'_f_TLL}
\end{align}
Note that there is no term proportional to $\sin^2\theta$ because 
$n_{j,\bm R}^x + n_{j+1,\bm R}^x$ is nonnegligible for frustration.
The operator $\mathcal A$ is given by
\begin{align}
 \mathcal A
 &= -\frac{\lambda H}{2v} \sum_{\braket{\bm R,\bm R'}}\int dx \, \bigl[
 \bigl\{(n_1^+n_2^z+n_1^zn_2^+)\sin(Hx/v) + (n_1^+\e_2+\e_1 n_2^+)\cos(Hx/v)\bigr\}\cos^2\theta 
 \notag \\
 &\quad
 - \bigl\{ (n_1^z n_2^z - \e_1 \e_2)\sin(2Hx/v) + (n_1^z \e_2 + \e_1 n_2^z)\cos(2Hx/v)
 \bigr\}\sin\theta\cos\theta
 \bigr],
\end{align}
which leads to
\begin{align}
  G^R_{\mathcal A\mathcal A^\dag}(\omega)
  &= \biggl(\frac{\lambda_\perp H}{2v}\biggr)^2 N\zeta
  \biggl[
  \cos^4\theta \bigl\{ G^R_{(\frac 12, \frac 12)}(\omega, H/v) + G^R_{(\frac 12, \frac 12)}(\omega,-H/v)\bigr\}
  \notag \\
  &\quad + \frac{\sin^2\theta\cos^2\theta}2 \bigl\{ G^R_{(\frac 12, \frac 12)}(\omega, 2H/v)
  +G^R_{(\frac 12, \frac 12)}(\omega,-2H/v)\bigr\}
  \biggr].
\end{align}
\end{widetext}
We obtain the linewidth,
\begin{equation}
 \eta = \frac{\pi^3\zeta \lambda_\perp^2 H^2}{32v^2T} (\cos^4\theta +\cos^2\theta).
  \label{width_f}
\end{equation}
The linewidth  induced by the frustrated interchain anisotropy
also becomes maximum at $\theta=0\mod \pi$ differently from the quadrupolar TLL.

\section{Summary}\label{sec:summary}

We proposed the method to detect the quasi-long-range spin-nematic order of the quadrupolar TLL phase in ESR measurements.
We showed that the linewidth of the paramagnetic resonance peak exhibits the unique angular dependence in the quadrupolar TLL phase
of the $S=1/2$ frustrated ferromagnetic chain system.
The characteristic angular dependence originates from the single fact 
that the transverse correlation $\braket{S_r^xS_0^x}$ is decaying  much 
faster than the longitudinal one $\braket{S_r^zS_0^z}$ and the nematic one $\braket{S_r^+S_{r+1}^+S_0^-S_1^-}$ at the wavevector $\bm q=0$.
Interestingly enough, many anisotropic interactions result in the same angular dependence 
of $\sin^2\theta\cos^2\theta$ [Fig.~\ref{width_comparison}~(b) and Table~\ref{table:width}] for the quadrupolar TLL.
In contrast, since both the transverse and longitudinal correlations decay algebraically in the standard TLL phase, 
the linewidth never shows the $\sin^2\theta\cos^2\theta$ dependence (Table~\ref{table:width}) at low magnetic fields.
We found that the linewidth of the standard TLL becomes maximum at a certain angle $\theta$ with the period $\pi$ at low magnetic fields.
This is in sharp contrast to the linewidth of the quadrupolar TLL which is maximized at $\theta=\pi/4 \mod \pi/2$ with the period $\pi/2$.
Our results relied crucially on two conditions: 
the temperature is lower than the excitation gap of an unpaired magnon [Eq.~\eqref{cond_T}]
and both the magnetization density and the magnetic field are smaller than the temperature [Eq.~\eqref{cond_M_relaxed}].
In the case of the standard TLL, the condition \eqref{cond_M_relaxed_TLL} similarly to Eq.~\eqref{cond_M_relaxed} of the quadrupolar TLL
is imposed.
Because of those conditions, our results (Table~\ref{table:width}) hold true in the low-field region
of the quadrupolar or standard TLL phase.

Our claim about the periodicity of the linewidth will be applicable to a spin nematic phase in higher dimensional systems in principle.
In addition, as we discussed in Sec.~\ref{sec:nematic_correlation}, 
the resonance frequency \eqref{w_r} is expected to give an order parameter of 
the long-range spin nematic order in higher dimensional systems,
which is discussed elsewhere~\cite{furuya_3dnematic}.

We used in this paper the Mori-Kawasaki approach and the Oshikawa-Affleck theory (or the self-energy approach) to discuss the linewidth.
The former is convenient for its wide scope of applicability if we accept the assumption that 
the paramagnetic peak has the single Lorentzian lineshape.
This is the most nontrivial assumption that we rely on.
On the other hand, the self-energy approach does not require such an ad hoc assumption although its application scope is rather limited.
We could extend the Oshikawa-Affleck theory, originally developed for intrachain interactions, 
in order to deal with interchain interactions.
This extension gave a justification of the single Lorentzian lineshape of the paramagnetic peak of the standard TLL
governed by interchain interactions, although only for the limited field direction.
In addition to those arguments of the linewidth to distinguish the quadrupolar TLL from the standard one,
we also gave in the appendix~\ref{app:uDM} an interesting example of the case where the self-energy approach is applicable but the Mori-Kawasaki approach is not.
It is the $S=1/2$ Heisenberg antiferromagnetic chain with the uniform Dzyaloshinskii-Moriya interaction.
It is an open problem to investigate effects of the uniform Dzyaloshinskii-Moriya interaction on ESR of the quadrupolar TLL.
The author hopes that the present paper will encourage experimental and theoretical studies on ESR of the quadrupolar TLL and,
more generally, ESR in the spin nematic phase.

\section*{Acknowledgments}

I am grateful to A. Furusaki, T. Giamarchi, T. Momoi, M. Oshikawa, M. Sato, E. Takata, and S. Takayoshi for illuminating discussions.
The present work is supported by JSPS KAKENHI Grants No. 16J04731.

\appendix

\section{Polarization dependence}\label{app:polarization}

Here we derive the relation \eqref{polarization} from an identity~\cite{OA_PRB},
\begin{equation}
 \mathcal G^R_{S^+S^-}(\omega)
  = \frac{2\langle S^z\rangle}{\omega-H} - \frac{\langle [\mathcal A, S^-]\rangle}{(\omega-H)^2}
  + \frac 1{(\omega-H)^2} \mathcal G^R_{\mathcal A\mathcal A^\dagger}(\omega),
  \label{identity_pm}
\end{equation}
with $\mathcal A\equiv [\mathcal H', S^+]$ and $\mathcal H'$ is the anisotropic interaction.
Integrating Eq.~\eqref{GR_SpSm} by parts twice, we can derive the identity \eqref{identity_pm} from equations of motion for $S^\pm$:
\begin{align}
 \frac{dS^+}{dt}&= -iHS^+ + i\mathcal A, \\
 \frac{dS^-}{dt}&= iHS^- - i\mathcal A^\dagger.
\end{align}
Likewise, we can obtain identities for $S^x$ and $S^y$.
Their equations of motion are
\begin{align}
 \frac{dS^x}{dt}&= HS^y -\mathcal A_{\rm i}, \\
 \frac{dS^y}{dt}&= -HS^x + \mathcal A_{\rm r},
\end{align}
where $\mathcal A_{\rm r}$ and $\mathcal A_{\rm i}$ are real and imaginary parts of $\mathcal A$.
Integrating by parts, we find a relation
\begin{align}
 \mathcal G^R_{S^xS^x}(\omega)
 &= \frac{H\langle S^z\rangle}{\omega^2-H^2} - \frac{i\omega}{\omega^2-H^2}\mathcal G^R_{\mathcal A_{\rm i}S^x}(\omega)
 \notag \\
 & \quad -\frac H{\omega^2-H^2}\mathcal G^R_{\mathcal A_{\rm r}S^x}(\omega). 
 \label{SxSx_pre}
\end{align}
The retarded Green's functions on the right hand side are subject to similar identities,
\begin{align}
 \mathcal G^R_{\mathcal A_{\rm i}S^x}(\omega)
 &= \frac{\omega}{\omega^2-H^2}\biggl(\langle [\mathcal A_{\rm i}, S^x]\rangle - \frac{iH}{\omega}\langle [\mathcal A_{\rm i}, S^y]\rangle\biggr)
 \notag \\
 & \quad - \frac H{\omega^2-H^2}\mathcal G^R_{\mathcal A_{\rm i}\mathcal A_{\rm r}}(\omega) + \frac{i\omega}{\omega^2-H^2}\mathcal G^R_{\mathcal A_{\rm i}\mathcal A_{\rm i}}(\omega), \\
 \mathcal G^R_{\mathcal A_{\rm r}S^x}(\omega)
  &= \frac{\omega}{\omega^2-H^2}\biggl(\langle [\mathcal A_{\rm r}, S^x]\rangle - \frac{iH}{\omega}\langle [\mathcal A_{\rm r}, S^y]\rangle\biggr)
 \notag \\
 & \quad - \frac H{\omega^2-H^2}\mathcal G^R_{\mathcal A_{\rm r}\mathcal A_{\rm r}}(\omega) + \frac{i\omega}{\omega^2-H^2}\mathcal G^R_{\mathcal A_{\rm r}\mathcal A_{\rm i}}(\omega).
\end{align}
When we restrict ourselves to the region $|\omega-H|\ll H$, those relations are greatly simplified to
\begin{align}
 \mathcal G^R_{\mathcal A_{\rm i}S^x}(\omega)
 &\approx \frac{\langle[\mathcal A_{\rm i}, S^-]\rangle}{2(\omega-H)} - \frac 1{2(\omega-H)} \mathcal G^R_{\mathcal A_{\rm i}\mathcal A^\dagger}(\omega),  
\end{align}
and
\begin{align}
 \mathcal G^R_{\mathcal A_{\rm r}S^x}(\omega)
 &\approx \frac{\langle[\mathcal A_{\rm r}, S^-]\rangle}{2(\omega-H)} - \frac 1{2(\omega-H)} \mathcal G^R_{\mathcal A_{\rm r}\mathcal A^\dagger}(\omega).
\end{align}
Furthermore, combining them with Eq.~\eqref{SxSx_pre}, we obtain
\begin{align}
 \mathcal G^R_{S^xS^x}(\omega)
 &\approx \frac{\langle S^z\rangle}{2(\omega-H)} - \frac{\langle [\mathcal A, S^-]\rangle}{4(\omega-H)^2}
 \notag \\
 & \quad
 + \frac1{4(\omega-H)^2}\mathcal G^R_{\mathcal A\mathcal A^\dagger}(\omega).
 \label{GR_SxSx}
\end{align}
Comparing Eqs.~\eqref{identity_pm} and \eqref{GR_SxSx}, we conclude
\begin{equation}
 \mathcal G^R_{S^xS^x}(\omega) \approx \frac 14\mathcal G^R_{S^+S^-}(\omega).
  \label{SxSx-Spm}
\end{equation}
It is straightforward to confirm a similar relation,
\begin{equation}
 \mathcal G^R_{S^yS^y}(\omega) \approx \frac 14 \mathcal G^R_{S^+S^-}(\omega).
\end{equation}

\section{Commutation relations}\label{app:ope}

Here we show a list of operator product expansion of $\bm J_R$, $\bm J_L$, $\bm n$ and $\e$
and also equal-time commutation relations derived from them.
On the flat two-dimensional plane, those operators satisfy operator product expansions~\cite{Starykh_checkerboard, Furukawa_chiral},
\begin{widetext}
\begin{align}
 \frac 1{\sqrt{8\pi^2}}J_{R/L}^a(z_{R/L}) J_{R/L}^b(0) &= \frac{\delta^{ab}}{\sqrt{8\pi^2} z_{R/L}^2} + \frac{if^{abc}J_{R/L}^c}{2\pi z_{R/L}},
 \\
 \frac 1{\sqrt{8\pi^2}}J_{R/L}^a(z_{R/L}) n^b(0) &= \frac i{4\pi z_{R/L}} \bigl[f^{abc}n^c(0) \pm \delta^{ab}\e(0)\bigr], \\
 \frac 1{\sqrt{8\pi^2}} J_{R/L}^a (z_{R/L}) \e(0) &= \frac{\mp i n^a(0)}{4\pi z_{R/L}},
\end{align}
where $z_R = v\tau-ix$ and $z_L=v\tau + ix$ are complex coordinates to describe the two-dimensional Euclidean spacetime
and
$f^{abc}$ is the three-dimensional completely antisymmetric tensor with $f^{xyz}=1$.
They can be rewritten in terms of equal-time commutation relations as~\cite{Gogolin_book}
\begin{align}
 \frac 1{\sqrt{8\pi^2}} [J_{R/L}^a(x), J_{R/L}^b(y)] &= \frac i{\sqrt{2\pi}} \delta^{ab}\partial_y\delta(x-y) + if^{abc} J_{R/L}^c(y)\delta(x-y),
 \label{comm_JJ_app} \\
 \frac 1{\sqrt{8\pi^2}} [J_{R/L}^a(x), n^b(y)] &= \frac i2  \bigl[ f^{abc} n^c(y) \pm \delta^{ab}\e(y)\bigr]\delta(x-y),
 \label{comm_Jn_app} \\
 \frac 1{\sqrt{8\pi^2}} [J_{R/L}^a(x), \e(y)] &= \mp \frac i2  n^a(y)\delta(x-y).
 \label{comm_Je_app}
\end{align}

\section{Chiral anomaly}\label{app:anomaly}

Now we discuss that the chiral anomaly has no impact on the linewidth of the paramagnetic peak.
That is, the first term of the commutation relation \eqref{comm_JJ} yields no contribution at $\omega=H$.
This term adds the following operators to $\mathcal A$:
\begin{align}
  &\frac{\lambda\sin^2\theta}2 \int dxdy \, \bigl( J_R^+ e^{-iHy/v} + J_L^+e^{iHy/v}\bigr)\partial_y \delta(x-y)
  \notag \\
  & \quad
  +\lambda\sin\theta \cos\theta \int dx dy \, \bigl( J_R^z e^{iH(x-y)/v} + J_L^z e^{iH(-x+y)/v}\bigr)\partial_y\delta(x-y)
  \notag \\
  &= \frac{iH\lambda\sin^2\theta}2 \int dx \, (J_R^+ e^{-iHx/v} - J_L^+ e^{iHx/v}) + iH\lambda \sin\theta \cos\theta \int dx \, (J_R^z - J_L^z).
\end{align}
\end{widetext}
Therefore, the chiral anomaly adds
\begin{equation}
 (H\lambda)^2\bigl[\sin^4\theta \, G^R_{(0,1)}(\omega,H/v)
  + 2\sin^2\theta \cos^2\theta \, G^R_{(0,1)}(\omega,0)\bigr],
\end{equation}
to the retarded Green's function $G^R_{\mathcal A\mathcal A^\dag}(\omega)$.
According to Eq.~\eqref{GR_vertex_generic_2}, 
the imaginary part of $G^R_{(0,1)}(\omega,q)$ is proportional to a delta function of $\delta(\omega+vq)$.
When $H/T\ll 1$, the chiral anomaly adds the following term to $\operatorname{Im}G^R_{\mathcal A\mathcal A^\dag}(\omega)$,
\begin{equation}
 \frac{\pi(H\lambda)^2}v \bigl[ \sin^4\theta (\omega-H)\delta(\omega+H) + \sin^2\theta\cos^2\theta \omega\delta(\omega)\bigr].
\end{equation}
It is  zero thanks to $\omega>0$.
In the end, we conclude that the chiral anomaly has no impact on the linewidth of the paramagnetic resonance peak at least at the second order of $\lambda$.

\section{Uniform DM interaction} \label{app:uDM}

This section is devoted to investigation of the linewidth of the standard TLL induced by the uniform DM interaction,
which is an example that can be dealt with by the self-energy approach but \emph{not} by the MK approach.
The model we consider here is the $S=1/2$ HAFM chain \eqref{H_HAFM_anisotropic} with a perturbative uniform DM interaction,
\begin{equation}
 \mathcal H' = \sum_j \bm D \cdot \bm S_j \times \bm S_{j+1}.
\end{equation}
We put the DM vector $\bm D$ on the $zx$ plane and rotate it so that $\bm D = D(\hat z \cos\theta + \hat x \sin \theta)$.
The total Hamiltonian is given by~\cite{Gangadharaiah_uDM}
\begin{align}
  \mathcal H
  &= \frac v{48\pi} \int dx \, (\bm J_R \cdot \bm J_R + \bm J_L \cdot \bm J_L) 
  \notag \\
  &\qquad - \frac H{\sqrt{8\pi^2}} \int dx \, (J_R^z +J_L^z)
  \notag \\
  &\quad + \frac{\gamma D\cos\theta}{\sqrt{8\pi^2}} \int dx \, (J_R^z - J_L^z) 
  \notag \\
  &\quad + \frac{\gamma D\sin\theta}{\sqrt{8\pi^2}}\int dx \, (J_R^x - J_L^x),
 \label{H_uDM}
\end{align}
where $\gamma$ is a nonuniversal constant.

\subsection{$\theta=0$}

Let us start with two simple cases of $\theta=0$ and $\pi/2$.
When $\theta=0$, the DM vector $\bm D$ is parallel to the magnetic field.
Interestingly, the right mover and the left mover feel different effective magnetic fields, $H_R^0$ and $H_L^0$.
In fact, the effective Hamiltonian \eqref{H_uDM} is written as
\begin{align}
  \mathcal H
  &= \frac v{48\pi} \int dx \, \bigl(\bm J_R \cdot \bm J_R + \bm J_L \cdot \bm J_L\bigr)
  \notag \\
  &\quad -\frac{H_R^0}{\sqrt{8\pi^2}} \int dx \, J_R^z - \frac{H_L^0}{\sqrt{8\pi^2}} \int dx \, J_L^z,
\end{align}
with 
\begin{equation}
 \left\{
  \begin{split}
   H_R^0 &= H - \gamma D, \\
   H_L^0 &= H + \gamma D.
  \end{split}
 \right.
\end{equation}
The chirality-dependent magnetic fields $H_{R/L}^0$ can be eliminated from the Hamiltonian
as we did in Eq.~\eqref{shift_phi}.
First we rewrite the Hamiltonian in terms of $\varphi_{R/L}$ as follows.
\begin{align}
 \mathcal H
 &= v\int dx \, \bigl\{(\partial_x\varphi_R)^2 + (\partial_x\varphi_L)^2\bigr\}
 \notag \\
 &\quad - \frac{2H_R^0}{\sqrt{2\pi}} \int dx \, \partial_x\varphi_R
 -\frac{2H_L^0}{\sqrt{2\pi}} \int dx \, \partial_x \varphi_L.
\end{align}
Next we eliminate $H_{R/L}^0$, shifting $\varphi_{R/L}$ by
\begin{equation}
 \left\{
  \begin{split}
   \varphi_R &\to \varphi_R + \frac 1{\sqrt{2\pi}} \frac{H_R^0 x}v, \\
   \varphi_L &\to \varphi_L + \frac 1{\sqrt{2\pi}} \frac{H_L^0 x}v.
  \end{split}
 \right.
 \label{shift_phi_chiral}
\end{equation}
The shift \eqref{shift_phi_chiral} modifies the bosonization formula of $S^\pm$ to
\begin{equation}
  S^\pm = \frac 1{\sqrt{8\pi^2}} \int dx \, (J_R^\pm e^{\pm iH_R^0x/v} + J_L^\pm e^{\mp iH_L^0x/v}),
  \label{Spm_J_uDM}
\end{equation}
where two wavenumbers $H_{R/L}^0/v$ emerges.
The retarded Green's function $G^R_{S^+S^-}(\omega)$ has two poles at $\omega= H_R^0$ and $\omega=H_L^0$:
\begin{align}
 G^R_{S^+S^-}(\omega)
 &= \frac N{4\pi^2} \biggl[ G^R_{(0,1)}(\omega, -H_R^0/v)
 + G^R_{(1,0)} (\omega, H_L^0/v)
 \biggr] 
 \notag \\
 &\approx \frac{NH}{2\pi v} \biggl(\frac 1{\omega - H_R^0 + i0}
 + \frac 1{\omega - H_L^0 + i0}
 \biggr).
\end{align}
In other words, the paramagnetic peak $\omega=H$ is split into $\omega=H_R^0$ and $\omega = H_L^0$.
The amount of the splitting $|H_R^0-H_L^0| = 2\gamma D$ is proportional to $D$.
The splitting is indeed observed in Ref.~\onlinecite{Povarov_ESR_DM}, which supports the argument given above.
Thus far the linewidth of the split peaks is yet to be discussed.
In the discussion given above, we induced the DM interaction up to the first order $(D/J)^1$.
A perturbative expansion of the linewidth starts from a higher order than $(D/J)^2$.
Thus we need to take a look at a correction of the Hamiltonian at higher order of $D/J$.
To do so, we go back to the lattice Hamiltonian
\begin{align}
  \mathcal H
  &= J \sum_j \bm S_j\cdot \bm S_{j+1} - HS^z 
 \notag \\
  &\quad + D\sum_j (S_j^xS_{j+1}^y - S_j^y S_{j+1}^x),
\end{align}
and perform a rotation,
\begin{equation}
 \begin{pmatrix}
  S_j^x \\ S_j^y \\ S_j^z 
 \end{pmatrix}
 =
 \begin{pmatrix}
  \cos(\alpha j) & - \sin(\alpha j) & 0 \\
  \sin(\alpha j) & \cos(\alpha j) & 0 \\
  0 & 0 & 1
 \end{pmatrix}
 \begin{pmatrix}
  \tilde S_j^x \\ \tilde S_j^y \\ \tilde S_j^z
 \end{pmatrix},
 \label{rot_uDM}
\end{equation}
with an angle $\alpha = \tan^{-1}(D/J)$.
The rotation eliminates the uniform DM interaction from the Hamiltonian 
and gives rise to an exchange anisotropy as a price for it,
\begin{align}
  \mathcal H
  &= \sqrt{J^2+D^2} \sum_j \tilde{\bm S}_j \cdot \tilde{\bm S}_{j+1}
  - HS^z 
  \notag \\
  &\quad
  +\bigl(J-\sqrt{J^2+D^2}\bigr)\sum_j \tilde S_j^z \tilde S_{j+1}^z.
\end{align}
At the lowest order of $D/J$, the exchange anisotropy is expressed as
\begin{equation}
 \mathcal H' \approx -\frac{D^2}{2J} \sum_j \tilde S_j^z\tilde S_{j+1}^z  \approx -\frac{D^2}{2J} \sum_j S_j^zS_{j+1}^z,
\end{equation}
or
\begin{equation}
 \mathcal H' \approx  - \frac{\Gamma D^2}{2J} \int dx \, J_R^z J_L^z,
\end{equation}
with a nonuniversal constant $\Gamma$.
Therefore, the uniform DM interaction with $\bm D=D\hat z$ induces the linewidth
\begin{equation}
 \eta = 4\pi^3\biggl(\frac{\Gamma D^2}{2J}\biggr)^2 \frac T{v^2},
\end{equation}
which is of the order of $(D/J)^4$.

\subsection{$\theta=\pi/2$}

When $\theta=\pi/2$, the DM vector $\bm D$ is perpendicular to the magnetic field.
Then the approach to eliminate the DM interaction by a rotation is not effective.
The rotation is to be performed around the $x$ axis and affects the Zeeman energy unpleasantly.
The Zeeman energy of the rotated system oscillates spatially with the wavenumber of $\tan^{-1}(D/J)$, 
which is difficult to be handled in the self-energy approach.
When $\theta=\pi/2$, the effective bosonized Hamiltonian is given by
\begin{align}
  \mathcal H
  &= \frac v{48\pi} \int dx \, (\bm J_R \cdot \bm J_R + \bm J_L \cdot \bm J_L) 
  \notag \\
  &\quad - \frac 1{\sqrt{8\pi^2}} \int dx \, (HJ_R^z - \gamma D J_R^x)
   \notag \\
  &\quad - \frac 1{\sqrt{8\pi^2}} \int dx \, (HJ_L^z + \gamma D J_L^x).
 \label{H_uDM_x}
\end{align}
Note that the right-moving part and the left-moving part are independent at the level of the effective Hamiltonian \eqref{H_uDM_x}.
This observation motivates us to rotate $\bm J_R$ and $\bm J_L$ differently so as to simplify the Hamiltonian.
Rotations
\begin{align}
 \begin{pmatrix}
  J_R^z \\ J_R^x
 \end{pmatrix}
 &=
 \begin{pmatrix}
  \cos \alpha_R & -\sin\alpha_R \\
  \sin\alpha_R & \cos\alpha_R
 \end{pmatrix}
 \begin{pmatrix}
  J'^z_R \\ J'^x_R
 \end{pmatrix},
 \label{rot_R} \\
 \begin{pmatrix}
  J_L^z \\ J_L^x
 \end{pmatrix}
 &=
 \begin{pmatrix}
  \cos \alpha_L & -\sin\alpha_L \\
  \sin\alpha_L & \cos\alpha_L
 \end{pmatrix}
 \begin{pmatrix}
  J'^z_L \\ J'^x_L
 \end{pmatrix},
 \label{rot_L}
\end{align}
with
\begin{equation}
 \alpha_R = -\alpha_L = \tan^{-1}\biggl(\frac{\gamma D}H\biggr).
   \label{alpha_R}
\end{equation}
The rotation simplifies the Hamiltonian to
\begin{align}
  \mathcal H 
  &= \frac v{48\pi} \int dx \, (\bm J'_R \cdot \bm J'_R + \bm J'_L\cdot \bm J'_L)
  \notag \\
  &\quad
  -\sqrt{\frac{H^2+\gamma^2 D^2}{8\pi^2}} \int dx \, (J'^z_R+J'^z_L).
\end{align}
In contrast to the $\theta=0$ case, the paramagnetic peak is not split.
However, this argument is incomplete
 because the decoupling of the right-moving and left-moving parts is imperfect in general.
The $S=1/2$ HAFM chain yields an isotropic interaction $g\bm J_R \cdot \bm J_L$ with $g\propto J$ in addition to Eq.~\eqref{H_uDM}.
This interaction was ignored thus far for its irrelevance in ESR~\cite{OA_PRB}.
In fact, it is marginally irrelevant in the RG sense and yields only a logarithmic correction to the resonance frequency and
the linewidth such as $\ln(J/T)$ in Eq.~\eqref{width_stag}.
However, since the interaction $g\bm J_R\cdot \bm J_L$ is variant under the chiral rotations \eqref{rot_R} and \eqref{rot_L},
we must take it into account.
The rotations modify it to
\begin{align}
 g\bm J_R \cdot \bm J_L
 &= g\cos(2\alpha_R)(J'^z_RJ'^z_L + J'^x_R J'^x_L) 
 \notag \\
 & \quad + g\sin(2\alpha_R) (J'^x_R J'^z_L - J'^z_R J'^x_L) + gJ'^y_R J'^y_L
 \notag \\
 &\approx g\bm J'_R \cdot \bm J'_L +\frac{2\gamma gD}H (J'^x_R J'^z_L - J'^z_R J'^x_L).
\end{align}
In the last line we have approximated it for small $D/H\ll 1$.
Thus, the chiral rotations \eqref{rot_R} and \eqref{rot_L} effectively generates the interaction,
\begin{equation}
 \mathcal H' = \frac{2\gamma gD}H \int dx \, (J'^x_RJ'^z_L - J'^z_RJ'^x_L).
  \label{H'_uDM_x}
\end{equation}
Hereafter we omit the prime for simplicity of notation.
The rotated Hamiltonian is thus
\begin{align}
  \mathcal H 
  &= \frac v{48\pi} \int dx \, (\bm J_R \cdot \bm J_R + \bm J_L\cdot \bm J_L)
  \notag \\
  &\quad
  -\sqrt{\frac{H^2+\gamma^2 D^2}{8\pi^2}} \int dx \, (J^z_R+J^z_L)
  \notag \\
  &\quad + \frac{2\gamma gD}H \int dx \, (J_R^xJ_L^z - J_R^z J_L^x),
\end{align}
where we dropped the isotropic irrelevant interaction again.
To investigate the linewidth, we evaluate the self-energy.
When $D/H\ll 1$, the retarded Green's function $\mathcal G^R_{S^+S^-}(\omega)$ is approximated as
\begin{align}
 \mathcal G^R_{S^+S^-}(\omega)
  &\approx \frac N{8\pi^2} \bigl[ \mathcal G^R_{J_R^x J_R^x}(\omega,-H')
  + \mathcal G^R_{J_L^x J_L^x}(\omega, H') 
  \notag \\
 &\quad + \mathcal G^R_{J_R^y J_R^y}(\omega,-H')
 + \mathcal G^R_{J_L^y J_L^y}(\omega, H')
  \bigr],
\label{GR_SpSm_uDM_x}
\end{align}
with $H'=\sqrt{H^2+\gamma^2 D^2}$.
As we did in Sec.~\ref{sec:intrachain}, we relate it to $\mathcal G^R_{\phi\phi}(\omega,q)$ by rotating the system by $\pi/2$ around a certain axis.
For example, the rotation $y \to z$ (the $\pi/2$ rotation around $\hat x$) changes $\mathcal G^R_{J_R^y J_R^y}(\omega,q)$ and $\mathcal H'$
into $4\pi(\omega-q)^2{\mathcal G^R_{\phi\phi}}^{y\to z}(\omega,q)$ and $\mathcal H'_{y\to z}$, respectively.
The rotated perturbation $\mathcal H'_{y\to z}$ is
\begin{align}
 \mathcal H'_{y\to z}
 &= \frac{2\gamma gD}H \int dx \, (J_R^x J_L^y - J_R^y J_L^x) 
 \notag \\
 &= \frac{4\gamma gD}H \int dx \, \cos\sqrt{8\pi}\phi.
 \label{H'_y2z_uDM_x}
\end{align}
Therefore, the self-energy $\Pi^R_{y\to z}$ of the retarded Green's function  ${\mathcal G^R_{\phi\phi}}^{y\to z}(\omega,q)$
is given by
\begin{equation}
 \Pi^R_{y\to z}(\omega,q) = 4\pi^2\biggl(\frac{4\gamma gD}H\biggr)^2 \bigl[ G^R_{(1,1)}(\omega,q) - G^R_{(1,1)}(0,0)\bigr].
  \label{Pi_y2z_uDM_x}
\end{equation}

Calculations of the Green's functions involved with $J_{R/L}^x$ are trickier.
First we perform the rotation $y\to z$ that changes $\mathcal G^R_{J_R^x J_R^x}(\omega,q)$ and $\mathcal H'$ into
${\mathcal G^R_{J_R^x J_R^x}}^{y\to z}(\omega,q)$ and Eq.~\eqref{H'_y2z_uDM_x}.
The anisotropy \eqref{H'_y2z_uDM_x} can also be written in terms of $n^z=\cos\sqrt{2\pi}\phi$ as
\begin{equation}
 \mathcal H'_{y\to z} = \frac{8\gamma gD}H \int dx \, (n^z)^2.
\end{equation}
Next we perform further rotation $z\to x$ so that the Green's function is changed to that of $J_R^z$, which we denote as
${\mathcal G^R_{J_R^zJ_R^z}}^{y\to z \to x}(\omega,q)= 4\pi(\omega-q)^2{\mathcal G^R_{\phi\phi}}_{y\to z\to x}(\omega,q)$,
and the perturbation is changed to
$\mathcal H'_{y\to z\to x} = (8\gamma D/H)\int dx \, (n^x)^2$, that is,
\begin{equation}
 \mathcal H'_{y\to z \to x} = \frac{4\gamma gD}H \int dx \, \cos\sqrt{8\pi}\tilde \phi.
  \label{H'_y2z2x_uDM_x}
\end{equation}
As it is discussed in Ref.~\onlinecite{OA_PRB}, we may identify $\mathcal G^R_{\phi\phi}$ and $\mathcal G^R_{\tilde\phi\tilde\phi}$ at the lowest order of the perturbation.
Finally, the problem is reduced to investigation of the self-energy $\Pi^R_{y\to z\to x}$ of 
the Green's function ${\mathcal G^R_{\tilde \phi\tilde\phi}}^{y\to z\to x}(\omega,q)$
under the perturbation \eqref{H'_y2z2x_uDM_x}, which obviously gives the same result with Eq.~\eqref{Pi_y2z_uDM_x}.
Combining all these results, we find that the Green's function $\mathcal G^R_{S^+S^-}(\omega)$ is approximated
at the lowest order of $D/H$ as
\begin{equation}
 \mathcal G^R_{S^+S^-}(\omega)
  = \frac{NH'}{2\pi} \frac 1{\omega - H' - \frac 1{2H'}\Pi^R_{y\to z}(H',H')}.
\end{equation}
The linewidth $\eta$ is thus given by
\begin{equation}
 \eta = 2\pi^2 \biggl(\frac{4\gamma gD}H\biggr)^2 T,
\end{equation}
which is quadratic in $D/J$.

\subsection{General angles}

\begin{figure}[b!]
 \centering
 \includegraphics[bb = 0 0 890 623, width=\linewidth]{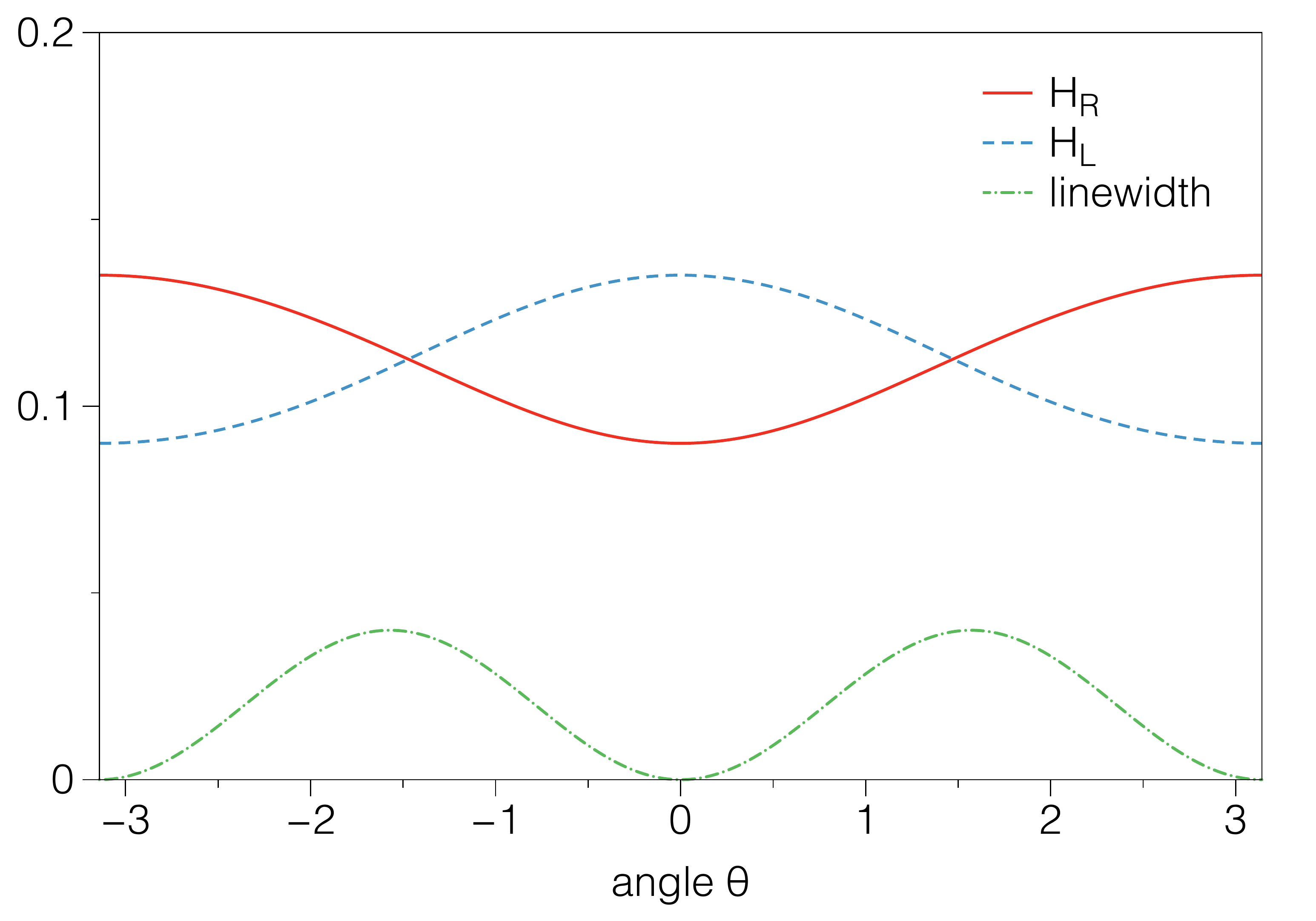}
 \caption{
 Angular dependence of the resonance frequencies \eqref{H_R} and \eqref{H_L} of the split paramagnetic resonance peaks
 and their linewidth \eqref{width_uDM} induced by the uniform DM interaction.
 Here we used a value $\gamma D/H=0.2$.
 }
 \label{fig:uDM}
\end{figure}

The results of two specific cases of $\theta=0$ and $\pi/2$ are easily extended to a case of general $\theta$.
When $D/H\ll 1$, 
chiral rotations \eqref{rot_R} with $\alpha_R = \tan^{-1}(\gamma D\sin\theta/(H-\gamma D\cos\theta))$ 
and \eqref{rot_L} with $\alpha_L = -\tan^{-1}(\gamma D\sin\theta/(H+\gamma D\cos\theta))$ transform the Hamiltonian \eqref{H_uDM} into
\begin{align}
  \mathcal H
  &= \frac v{48\pi} \int dx \, (\bm J_R \cdot \bm J_R + \bm J_L \cdot \bm J_L)
  \notag \\
  &\quad - \int dx \, (H_R J_R^z + H_L J_L^z)
  \notag \\
  &\quad + \frac{4\gamma gD\sin\theta}H \int dx \, (J_R^x J_L^z - J_R^z J_L^x),
\end{align}
with effective magnetic fields,
\begin{align}
 H_R &= \sqrt{\frac{(H-\gamma D\cos\theta)^2 + (\gamma D\sin\theta)^2}{8\pi^2}},
 \label{H_R}\\
 H_L &= \sqrt{\frac{(H+\gamma D\cos\theta)^2 - (\gamma D\sin\theta)^2}{8\pi^2}}.
 \label{H_L}
\end{align}
Note that Eqs.~\eqref{H_R} and \eqref{H_L} are valid for $D/H\ll 1$.
The Green's function $G^R_{S^+S^-}(\omega)$ have two Lorentzian peaks with a finite linewidth,
\begin{align}
 \mathcal G^R_{S^+S^-}(\omega)
 &= \frac N{8\pi^2} \bigl[\mathcal G^R_{J_R^xJ_R^x}(\omega, -H_R) 
 + \mathcal G^R_{J_L^x J_L^x}(\omega, H_L) \notag \\
 & \quad + \mathcal G^R_{J_R^y J_R^y}(\omega, -H_R) 
 + \mathcal G^R_{J_L^y J_L^y}(\omega, H_L)
 \bigr]
 \notag \\
 &\approx \frac{NH_R}{2\pi} \frac 1{\omega - H_R - \frac 1{2H}\Pi^R(H,H)}
 \notag \\
 & \quad + \frac{NH_L}{2\pi} \frac 1{\omega - H_L - \frac 1{2H}\Pi^R(H,H)}.
\end{align}
The self-energy is given by $\Pi^R(\omega,q)=4\pi^2(4\gamma D\sin\theta/H)^2[G^R_{(1,1)}(\omega,q)-G^R_{(1,1)}(0,0)]$,
which leads to the linewidth,
\begin{equation}
 \eta =2\pi^2\biggl(\frac{4\gamma D\sin\theta}H\biggr)^2 T.
  \label{width_uDM}
\end{equation}
The angular dependence $\eta \propto \sin^2\theta$ has the periodicity $\pi$ (Fig.~\ref{fig:uDM}).

\end{document}